\documentclass{osa-article}
\journal{oe}
\usepackage{mathtools}
\DeclareCaptionFormat{myformat}{#1#2#3\hrulefill}
\usepackage{subcaption}
\usepackage{url} 
\usepackage{bm} 
\DeclareMathOperator*{\argmin}{arg\,min}

\articletype{Research Article}

\usepackage{lineno}

\begin{document}

\title{Subspace tracking for phase noise source separation in frequency combs}

\author{Aleksandr Razumov,\authormark{1} 
 Holger R.~Heebøll,\authormark{1}, Mario Dummont\authormark{2}, Osama Terra \authormark{2,4},
 Bozhang Dong \authormark{2},
 Jasper Riebesehl \authormark{1},
 Poul Varming \authormark{3},
 Jens E.~Pedersen \authormark{3},
 Francesco Da Ros \authormark{1},
 John E.~Bowers \authormark{2} and
 Darko Zibar \authormark{1,*}}

\address{\authormark{1}Department of Electrical and Photonics Engineering, Technical University of Denmark (DTU), DK-2800 Kgs. Lyngby, Denmark\\}
\address{\authormark{2}Department of Electrical and Computer Engineering, University of California, Santa Barbara, CA 93106, USA\\}
\address{\authormark{3}NKT Photonics, Blokken 84, DK-3460 Birkerød, Denmark\\}
\address{\authormark{4}Primary length and laser technology lab, National Institute of Standards, Giza, Egypt\\}

\email{\authormark{*}dazi@dtu.dk} 






\begin{abstract} 
It is widely acknowledged that the phase noise of an optical frequency comb primarily stems from the common mode (carrier-envelope) and the repetition rate phase noise. However, owing to technical noise sources or other intricate intra-cavity factors, residual phase noise components, distinct from the common mode and the repetition rate phase noise, may also exist. We introduce a measurement technique that combines subspace tracking and multi-heterodyne coherent detection for the separation of different phase noise sources. This method allows us to break down the overall phase noise sources associated with a specific comb-line into distinct phase noise components associated with the common mode, the repetition rate and the residual phase noise terms. The measurement method allow us, for the first time, to identify and measure residual phase noise sources of a frequency modulated mode-locked laser.
\end{abstract}

\section{Introduction} \label{sec:intro}
The performance of optical frequency combs is determined by the noise of the comb-lines \cite{Diddams00,Diddams2020,Fortier20,Newbury:07}. Phase noise sets a limit on the sensitivity, information, and secret key rates for applications related to metrology, sensing and communication \cite{Diddams2020,Hussein:15,Lundberg2020}. Measuring, understanding, and reducing phase noise is therefore of great importance. \vspace{0.1cm}

In a theoretically perfect comb, there are two main sources of phase noise: 1) the carrier envelope (common mode) phase noise, $\phi^{cm}(t)$, that is common to all comb-lines, and 2) the repetition rate phase noise, $\phi^{rep}(t)$, (timing jitter), that increases linearly with comb-line number $m$ \cite{Newbury:07,Paschotta06,Fortier20}. This implies that the phase noise of the $m$th comb-line is fully described as: $ \phi^m(t) = \phi^{cm}(t) + m\phi^{rep}(t)$. Complete phase noise characterization of an optical frequency comb therefore requires measurement of $\phi^{cm}(t)$ and $\phi^{rep}(t)$, as well as their scaling as a function of comb-line number $m$. \vspace{0.1cm}

In a real-world system, fluctuations originating from the technical noise or some other internal or external sources of noise that are not easily describe mathematically, may contribute to the total phase noise of the comb-line $m$ \cite{Benkler:05,Paschotta06,Tian21}. We denote this contribution as residual phase noise term $\phi^{x}(m,t)$. The total phase noise associated with a comb-line $m$ is then expressed as: $\phi^m(t)  =  \phi^{cm}(t) + m\phi^{rep}(t) + \phi^{x}(m,t)$. Since the physical process behind the residual phase noise term can be quite complex and unknown, the scaling of $\phi^{x}(m,t)$ with comb-line number $m$ is typically not known. Finally, there may potentially be several different sources of residual phase noise that scale differently with comb line number, ~i.e.~$\phi^x(m,t)=\phi_1^x(m,t)+\phi_2^x(m,t)...$. Even though the residual sources of phase noise may be low in magnitude, they potentially scale non-linearly such that their contributions in comb-lines far away from the central frequency becomes significant. This is specially important in applications like supercontinuum generation where  nonlinear effects are used to broaden the initial comb \cite{Obrzud:19,Song22,Cai:23}.

For real-world optical frequency combs, full phase noise characterization is thus achieved by measuring all phase noise terms,$[\phi^{cm}(t), \phi^{rep}(m,t), \phi_1^x(m,t), \phi_2^x(m,t),...]$ and their corresponding scaling as a function of comb-line number. The objective of this paper is to propose a measurement method able to identify and separate different phase noise contributions, and identify their corresponding scaling, using a single measurement set-up relying on multi-frequency heterodyne detection and digital signal processing.\vspace{0.1cm}

Various techniques have been reported in the literature on how to measure carrier envelope offset (common mode), and repetition rate phase noise \cite{Nishimoto:20,Vedala2017,Casanova:20,Roos:04}. Measurement techniques in \cite{Nishimoto:20,Vedala2017,Casanova:20,Roos:04} neglect the presence of residual phase noise terms, even though they may potentially have an impact on the outcome of the measurement. If residual phase noise sources are present, not identifying them, and stabilizing them, would limit the achievable time and frequency stability of the frequency comb. Moreover, magnitude of  residual phase noise terms can be used to benchmark the stability of the comb.\vspace{0.1cm}

In a paper by Haus and Mecozzi \cite{Haus:93}, it is shown that noise correlation can be used to identify  various sources of noise. In \cite{Ansquer21}, a spectrally resolved multipixel homodyne detection method was used to measure the correlation matrices for amplitude and phase noises of multiple optical comb-lines for a mode-locked Ti-sapphire laser. However, the employed measurement setup is highly complex, requiring various feedback stabilization schemes, and it's not clear how to distinguish measurement noise from the frequency comb phase noise sources. Additionally, a significant issue with the method is that the employed eigendecomposition produces noise modes that are scaled, making it difficult to determine the correct magnitudes of the power spectral densities of various phase noise sources. This same issue is also present in \cite{Brajato2020}. Finally, the references \cite{Ansquer21,Brajato2020}, do not provide experimental evidence that the measured phase noise sources are indeed correct nor do they address measurement of the residual phase noise terms.  
Thus, a solution that can identify and measure residual phase noise sources, and also accurately determine the true values of the corresponding power spectral densities and overcome the scaling issue would be valuable in practice.\vspace{0.1cm}

To the best of our knowledge, results presented in \cite{Benkler:05} are the only attempt to measure the residual phase noise term. However, the measurement set-up presented in \cite{Benkler:05} is highly complex requiring different set-ups, and measures the \textit{total} residual phase noise contribution. The measurement of scaling of the residual phase noise term is also not reported.\vspace{0.1cm}

Generally speaking, the problem of identifying and separating phase noise sources associated with comb-lines falls under the category of blind source separation, and can be found in various areas of engineering. To perform the identification and separation of phase noise sources, it is necessary to jointly process phase noise of multiple comb-lines and utilize advanced digital signal processing (DSP) techniques to perform phase noise source separation. Fortunately, the latest advancements in technology have provided us with state-of-the-art, balanced receivers and real-time sampling oscilloscopes (analog-to-digital converters) that have surpassed analog bandwidths of 100 GHz \cite{keysightoscilloscopes}. This breakthrough is opening up new opportunities for joint detection of a large number of comb-lines.\vspace{0.1cm}

In this paper, we propose a novel framework for phase noise source separation using multi-heterodyne detection and subspace tracking. We present the mathematical formalism and a scaling procedure that is essential for accurately computing the power spectral densities of the separated phase noise sources. We demonstrate that is possible to recover the common mode, the repetition rate and the residual phase noise sources, and their corresponding scaling, by jointly processing multiple-comb lines using subspace tracking. We show that the eigenvectors of the phase noise correlation matrix are directly related to the scaling of phase noise sources as a function of comb-line number.
The presented method assumes that the phase noise sources add linearly and have different scaling with comb-line number, i.e.~linearly independent phase noise sources. 

We demonstrate the effectiveness of our method by performing phase noise characterization of and electro-optic comb and a frequency modulated (FM) mode-locked quantum dot semiconductor laser-based comb.  Our results show that we can successfully identify and accurately measure independent sources of phase noise that contribute to the overall phase noise performance.\vspace{0.15cm}


The paper is organized as follows: In Section \ref{sec:theory}, we present the theoretical framework for subspace tracking aimed at identifying and separating phase noise sources. In Section \ref{sec:num_results}, we present detailed numerical simulations for multi-heterodyne coherent detection employing subspace tracking, demonstrating accurate phase noise source identification and estimation. In Section \ref{sec:exp_results}, we demonstrate experimental results using multi-heterodyne detection in combination with subspace tracking for phase noise source identification and measurement of electro-optic comb and FM mode-locked laser. Finally, in Section \ref{sec:conc}, we summarize the main findings and discuss the impact of our results.

\section{Theoretical framework}\label{sec:theory}

\begin{figure}[t!]
\centering \includegraphics[width=0.8\textwidth]{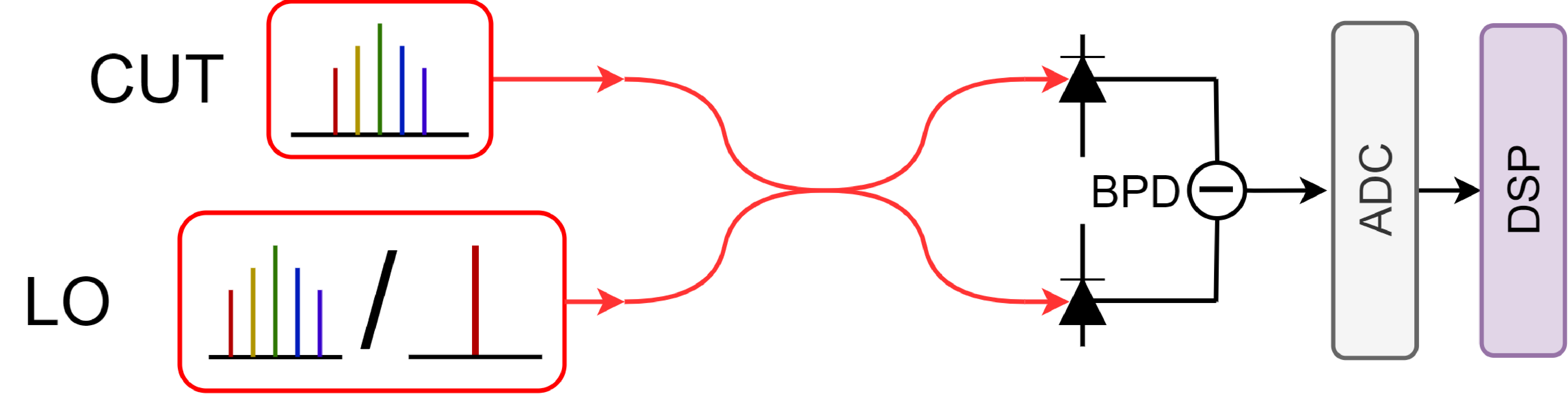}
\caption{Multi-heterodyne balanced detection for simultaneous detection of $M$ different frequency comb-lines. BPD: balanced photodetection, ADC: analogues-to-digital conversion. DSP: digital signal processing}
\label{fig:multi_het_setup}
\end{figure}

We assume a multi-heterodyne balanced detection followed by an analogue-to-digital converter (ADC), see Fig.~\ref{fig:multi_het_setup}. By multi-heterodyne signal detection, it is implied that the optical frequency comb under test (CUT) is combined with either a local oscillator (LO) single-frequency laser or a LO optical frequency comb in a balanced receiver. We assume an ideal LO with zero phase noise. The detected, and sampled, beat signal after the ADC, which only contains phase noise contributions from the CUT, becomes then:

\begin{equation}\label{eq:signal_coh_det}
    y_k = \sum_{m=-(M{-}1)/2}^{(M{-}1)/2} a_m\cos[2\pi (f_{0} + m\Delta f) kT_s + \phi^m_k] + n_k
\end{equation}

where $k$ is a discrete time index and $T_s$ is sampling time inversely proportional to the sampling frequency, $F_s$, of the ADC, i.e.~($T_s=1/F_s$). $M$ represents the total number of detected CUT comb-lines. Without loosing generality, we assume it is an odd number. The offset between the central frequency of CUT and LO is denoted by $f_0$, $\Delta f$ is a frequency spacing between the detected beat signal comb-lines, $a_m$ is the amplitude of a frequency component $m$, $\phi^m_k$ is the phase noise associated with comb-line $m$ and $n_k$ is an additive noise terms with Gaussian distribution. It has a zero mean and a variance, $\sigma^2_n$.

It has been shown in that the impairments associated with the ADC, such as limited resolution, noise and jitter, will manifest itself as a reduction in a signal-to-noise ratio \cite{Towfic10,Laperle14}. This implies that the impairments associated with the ADC can be modelled as an additive white Gaussian noise \cite{Towfic10,Laperle14}. Therefore, the noise term $n_k$, and the corresponding noise variance, $\sigma^2_n$, include contributions from thermal and shot noise \cite{Zibar2021}, as well as impairments associated with the ADC. \vspace{0.1cm}

In some cases, the output of CUT will be optically amplified. It has been shown in \cite{Zibar2021} that the impact of optical amplification can be modelled as an additive white Gaussian noise. This implies that the noise term $n_k$ can also encompass the impact of optical amplifier noise. Throughout this paper, we will refer to $n_k$ as measurement noise. \vspace{0.1cm}

In this paper, we would like to identify noise sources associated with $\phi^m_k$, and not $n_k$. We start by expression the phase noise of the m$th$ comb-line, at a discrete time instant $k$, in a mathematical form:

\begin{eqnarray}\label{eq:elastic tape model gen}
 \phi^m_k  & = &\phi^{cm}_k + m\phi^{rep}_k +\phi^{x}(m,k) \\ \nonumber  
        & = & \phi^{cm}_k + m\phi^{rep}_k + \overbrace{\phi_1^{x}(m,k)+\phi_2^{x}(m,k) + \cdots+\phi_P^{x}(m,k)}^{\phi^{x}(m,k)} \\  \nonumber
        & \equiv & \phi^{s_1}_k + m\phi^{s_2}_k + a(m)\phi^{s_3}_k +b(m)\phi^{s_4}_k+\cdots+ p(m)\phi^{s_P}_k \\ \nonumber 
    \end{eqnarray}

Finally, the phase noise of the m$th$ comb-line, $\phi^m_k$, can be expressed as a vector product between the vector containing scaling factors and the vector containing phase noise sources:  

\begin{eqnarray}\label{eq:elastic tape model gen vec}
 \phi^m_k =
    \begin{bmatrix} 
    1 & m & a(m) & b(m) & \hdots & p(m)
    \end{bmatrix}
     \begin{bmatrix}
    \phi^{s_1}_k \\
    \phi^{s_2}_k \\
    \phi^{s_3}_k \\
    \phi^{s_4}_k \\
    \vdots \\
    \phi^{s_P}_k
    \end{bmatrix} 
 \end{eqnarray}

where we have assumes that the residual phase noise term $\phi^{x}(m,k)$ contains $P$ different terms, each with a different scaling with $m$, and $s_1,...,s_P$ are indices associated with the number of phase noise sources. Since the scaling of residual phase noise terms, as a function of $m$, is not known it is denoted by the functions $a(m)$, $b(m)$ and $p(m)$. We assume that the phase noise sources add linearly to the $m$th comb-line phase noise $\phi^m_k$. This implies that $m\neq a(m) \neq b(m) \neq p(m)$. The objective is to identify all phase noise terms that add linearly, $[\phi^{s_1}_k,...,\phi^{s_P}_k]$, and determine their scaling as a function of comb-line number. To achieve that, we first need to simultaneously estimate phase noise $\phi^m_k$ of $M$ different comb-lines. The number of comb-lines $M$ needs to be greater than the anticipated number of phase noise source $P$. \vspace{0.1cm}

To estimate the phase noise, $\phi^m_k$, of $M$ different comb-lines, ideal bandpass filtering centered around $f_0 + m\Delta f$, with a bandwidth $B$, is performed, followed by the Hilbert transform to obtain quadrature signal, see Fig.~\ref{fig:multi_het_setup}. After that, we employ phase estimation based on an argument of the signal as expressed by Eq.~(9) in reference \cite{Zibar2021}. Assuming that the measurement noise is negligible, $n_k\approx 0$, a discrete time-series vector containing phase noise components is obtained: $\bm{\phi}_k=[\phi^{-(M{-}1)/2}_k,...,\phi^{(M-1)/2}_k]^T$, where $T$ is a transpose operator and $k=1,..,K$, where $K$ is the total number of discrete time samples. 


Assuming that there are $P$ sources of phase noise, the $M$-dimensional vector $\bm{\phi}_k$, at the discrete time instant $k$, can be decomposed into a product of a (generation) matrix $\mathbf{H}$ of $M\times P$ dimension and a vector of phase noise sources $\bm{\phi}^s_k=[\phi^{s_0}_k,...,\phi^{s_P}_k]^T$ of dimension $P$ as according to Eq.~(\ref{eq:elastic tape model gen}). We will make an assumption that $P<M$, which indeed can be satisfied in most practical cases. Using \eqref{eq:elastic tape model gen vec}, phase noise of the detected comb-lines, starting from $-(M-1)/2$ and ending with $(M-1)/2$ line, can expressed as:

\begin{eqnarray}\label{eq:forward model matrix}
    \begin{bmatrix}
    \phi^{-(M{-}1)/2}_k \\
    \vdots \\
    \phi^{(M{-}1)/2}_k
    \end{bmatrix} &=&
\begin{bmatrix}
    1 & -(M-1)/2 & a{(-(M-1)/2)} & \hdots & p{(-(M-1)/2)}\\
    \vdots & \vdots & \vdots & & \vdots \\
    1 & (M-1)/2 & a{((M-1)/2)} & \hdots & p{((M-1)/2)}\\
    \end{bmatrix}
    \begin{bmatrix}
    \phi^{s_1}_k \\
    \phi^{s_2}_k \\
    \phi^{s_3}_k \\
    \vdots \\
    \phi^{s_P}_k
    \end{bmatrix} \\  \nonumber
   & \equiv &
    \overbrace{ \begin{bmatrix}
    h_{11} & \hdots & h_{1P}\\
    \vdots & \ddots & \vdots \\
    h_{M1} & \hdots & h_{MP}\\
    \end{bmatrix}}^{\mathbf{H}}
    \begin{bmatrix}
    \phi^{s_1}_k \\
    \vdots \\
    \phi^{s_P}_k
    \end{bmatrix}
\end{eqnarray}

To keep things as general as possible, we assume that coefficients of matrix $\mathbf{H}$ are unknown. In a more compact notation \eqref{eq:forward model matrix} can be expressed as:

\begin{equation}\label{eq:forward model compact}
    \bm{\phi}_k = [\mathbf{h}_1,...,\mathbf{h}_P]\begin{bmatrix}
    \phi^{s_1}_k \\
    \vdots \\
    \phi^{s_P}_k
    \end{bmatrix}
    =\mathbf{H}\bm{\phi}^s_k
\end{equation}

where the columns of $\mathbf{H}=[\mathbf{h}_1,...,\mathbf{h}_p]$ are assumed to be linearly independent. According to \eqref{eq:forward model compact}, the matrix $\mathbf{H}$ describes how independent phase noise sources $\bm{\phi}^s_k$ are mapped to the total phase noise of a comb line $m$. The rank of the matrix $\mathbf{H}$ corresponds to the number of linearly independent columns of $\mathbf{H}$, and thereby to the number of linearly independent sources of phase noise.  

When $P<M$, the set of all possible signal vectors $\bm{\phi}_k$ lie in the subspace of the Euclidean space $\mathbb{R}^{M}$ spanned by the columns of $\mathbf{H}$. This subspace $\mathbb{R}^P$ is referred to as signal subspace. The phase noise vector $\bm{\phi}^s_k$ represents $P$ independent phase noise sources that add linearly to the total phase noise $\bm{\phi}_k$. In this paper, we refer to subspace tracking as the ability to identify the dimensionality of the phase noise sources (to find the correct subspace), and perform their idenitification and separation.  

\eqref{eq:forward model matrix} represents a time-domain description of phase noise generation process. An equivalent frequency domain description can also be derived.  

A common approach for characterizing phase noise of optical frequency combs is to compute phase noise power spectral density (PSD) associated with each comb-line. The autocorrelation function of the $m^{th}$ comb-line can be expressed as:

\begin{equation}\label{eq:AC}
    \gamma^m_{\phi\phi}(l) = E[\phi^m_k\phi^m_{k+l}]
    = h^2_{a1}E[\phi^{s_1}_k\phi^{s_1}_{k+l}]
     +...+ h^2_{aP}E[\phi^{s_P}_k\phi^{s_P}_{k+l}]
\end{equation}

where $l$ is an integer and $1\leqslant a \leqslant M$. The PSD of $m^{th}$ phase noise component is obtained by taking the Fourier transform of autocorrelation function $\gamma^m_{\phi\phi}(l)$: 

\begin{equation}\label{eq:PSD matrix}
    \begin{bmatrix}
        \Gamma^{-(M{-}1)/2}(f)\\
        \vdots \\
        \Gamma^{(M{-}-1)/2}(f)\\
    \end{bmatrix}
    = \begin{bmatrix}
        h^2_{11} & \hdots & h^2_{1P} \\
        \vdots & \ddots & \vdots \\
        h^2_{M1} & \hdots & h^2_{MP} \\
    \end{bmatrix}
\begin{bmatrix}
        \Gamma^{s_1}(f)\\
        \vdots \\
        \Gamma^{s_P}(f)\\
    \end{bmatrix}
\end{equation}

where $[\Gamma^{(-(M{-}1)/2)},...,\Gamma^{(M{-}1)/2}(f)]$ expresses the overall phase noise PSD of each comb line and $[\Gamma^{s_1}(f),...,\Gamma^{s_P}(f)]$ expresses PSD of the phase noise sources. Rows of the generation matrix $\mathbf{H}$ express how PSD of independent phase noise sources is added to the overall PSD of combs lines. \vspace{0.2cm}

Next, we would like to separate and estimate the independent phase noise sources $\bm{\phi}^s_k$ as they can be considered to be fundamental noise sources contributing to the overall phase noise of each comb line.  This is achieved by finding a projection matrix $\mathbf{G}=[\mathbf{g}_1,...,\mathbf{g}_P]$ with dimensions $P\times M$ that can be used to recover the  original phase noise sources $\bm{\phi}^{s}_k$ by a matrix multiplication:

\begin{equation}\label{eq:inverse model matrix}
    \begin{bmatrix}
    \phi^{s_1}_k \\
    \vdots \\
    \phi^{s_P}_k
    \end{bmatrix}=
     \begin{bmatrix}
    g_{11} & \hdots & g_{1M}\\
    \vdots & \ddots & \vdots \\
    g_{P1} & \hdots & g_{PM}\\
    \end{bmatrix}
    \begin{bmatrix}
    \phi^{-(M{-}1)/2}_k \\
    \vdots \\
    \phi^{(M{-}1)/2}_k
    \end{bmatrix}
\end{equation}

which can be expressed in more compact notation as:

\begin{equation}\label{eq:inverse model compact}
    \bm{\phi}^s_k = [\mathbf{g}_1,...,\mathbf{g}_P]^T
    \begin{bmatrix}
    \phi^{-(M{-}1)/2}_k \\
    \vdots \\
    \phi^{(M{-}1)/2}_k
    \end{bmatrix}
    = \mathbf{G}\bm{\phi}_k
\end{equation}

The rows of $\mathbf{G}$ form a basis of the $P-$dimensional subspace in which the phase noise sources, $\bm{\phi}^s_k\in \mathbb{R}^P$ exist. In the case when the generation matrix $\mathbf{H}$ is known and has linearly independent columns (has full rank), the projection matrix $\mathbf{G}$ can be computed using Pseudo-Moore inverse \cite{Bishop2006}:

\begin{equation}\label{eq:pseudoInverse}
    \mathbf{G}=(\mathbf{H}^T\mathbf{H})^{-1}\mathbf{H}^T
\end{equation}

which implies that:

\begin{equation}\label{eq:LS_solution}
    \bm{\phi}^s_k=(\mathbf{H}^T\mathbf{H})^{-1}\mathbf{H}^T\bm{\phi}_k
\end{equation}

For instance, if phase noise associated with each comb-line $m$ only contains contributions from the common mode and repetition rate phase noise, for that particular case \eqref{eq:forward model matrix} reduces to:

\begin{equation}\label{eq:forward model matrix firsr order}
    \begin{bmatrix}
    \phi^{-(M-1)/2}_k \\
    \vdots \\
    \phi^{(M-1)/2}_k
    \end{bmatrix}=
    \overbrace{\begin{bmatrix}
    1 & -(M{-}1)/2\\
    \vdots & \vdots \\
    1 & (M{-}1)/2 \\
    \end{bmatrix}}^{\mathbf{H}}
    \begin{bmatrix}
    \phi^{s_1}_k \\
    \phi^{s_2}_k
    \end{bmatrix}
    = 
    \begin{bmatrix}
    \mathbf{h}_1 & \mathbf{h}_2\\
    \end{bmatrix}
    \begin{bmatrix}
    \phi^{s_1}_k \\
    \phi^{s_2}_k
    \end{bmatrix}
\end{equation}

Employing Pseudo-Moore inverse, Eq.~(\ref{eq:pseudoInverse}), the following projection matrix, $\mathbf{G}$, is obtained:

\begin{equation}\label{eq:matrixG_first_order}
    \begin{bmatrix}
    \phi^{s_1}_k \\
    \phi^{s_2}_k
    \end{bmatrix}=
     \begin{bmatrix}
   \mathbf{h}^T_1/||\mathbf{h}_1||^2 \\
    \mathbf{h}^T_2/||\mathbf{h}_2||^2 \\
\end{bmatrix}
\begin{bmatrix}
    \phi^{-(M-1)/2}_k \\
    \vdots \\
    \phi^{(M-1)/2}_k
    \end{bmatrix} 
\end{equation}

Finally, common mode and repetition rate phase noise could originate from a sum of different phase noise sources, i.e.~$\phi^{s_1}_k = \phi^{s_{11}}_k + \phi^{s_{12}}_k+...+$. For that particular case, only the sum of them, $\phi^{s_1}_k$, can estimated due to the linear dependence.

In many practical cases, the exact size and coefficients of the generation matrix $\mathbf{H}$ are unknown. In that particular case, given the phase noise vector $\bm{\phi}_k$, finding the projection matrix $\mathbf{G}$ can be formulated in terms of the minimization of the average reconstruction error, $J(\mathbf{\tilde{G}})$, defined as \cite{murphy2013machine}:
\begin{equation}\label{eq:min J}
    \mathbf{G} = \argmin_{\tilde{\mathbf{G}}} \{J(\tilde{\mathbf{G}})\} 
\end{equation}

where $\mathbf{\tilde{G}}$ are trial projection matrices over which the optimization performed. The average reconstruction error is defined as:

\begin{equation}\label{eq:reconstruction error}
    J(\tilde{\mathbf{G}}) =  \frac{1}{K} \sum_{k=1}^K \parallel \bm{\phi}_k - \tilde{\bm{\phi}_k} \parallel^2
    =  \parallel \bm{\phi}_k - \mathbf{\tilde{G}}\mathbf{\tilde{G}}^T\bm{\phi}_k \parallel^2
\end{equation}

\begin{figure}[h!]
\centering \includegraphics[width=0.45\linewidth]{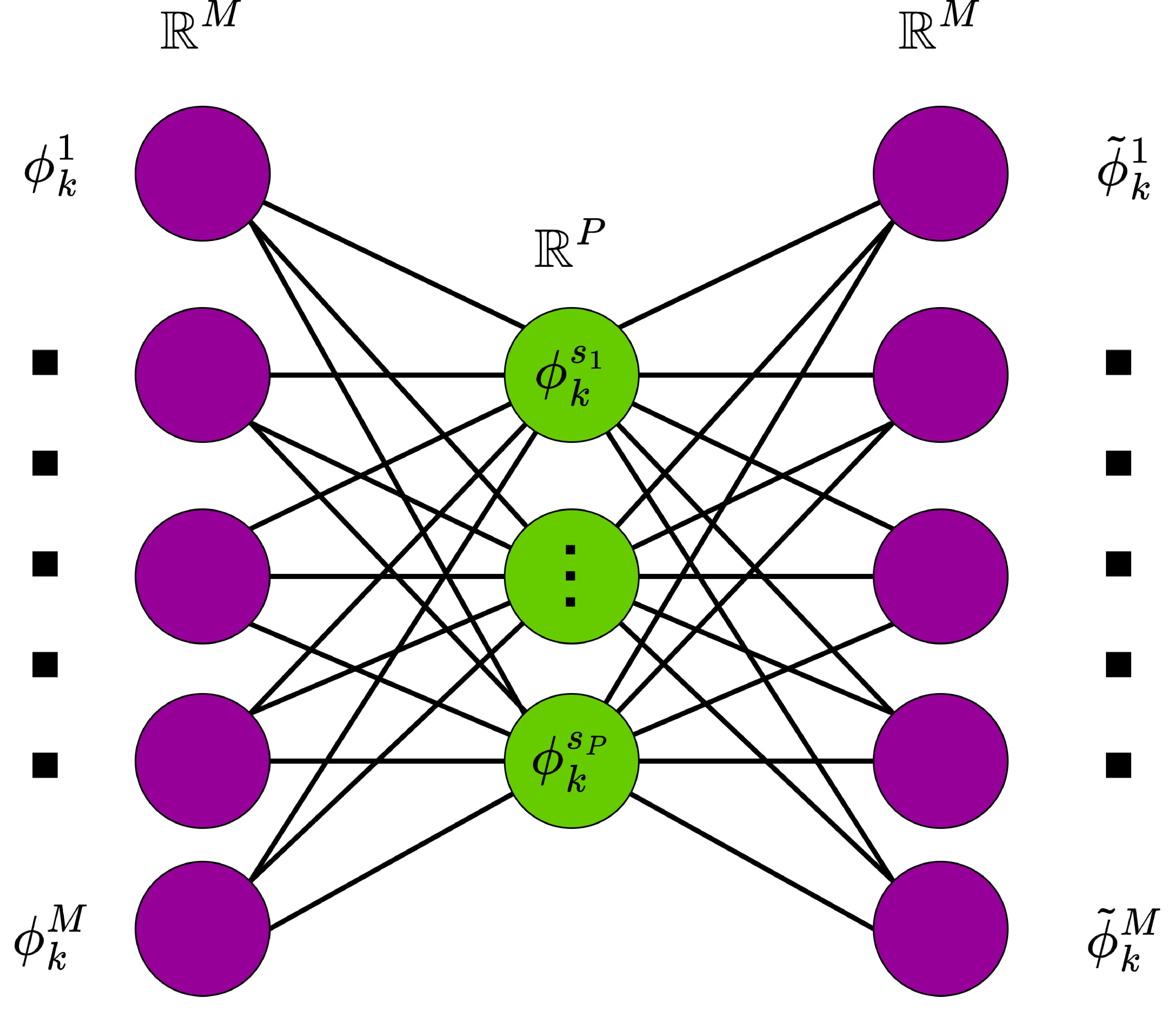}
\caption{Graphical illustration of the concept behind subspace tracking.}
\label{autoencoder}
\end{figure}

When the matrix $\mathbf{\tilde{G}}$ that solves the minimization problem in \eqref{eq:min J} has been found, we have $\mathbf{\tilde{G}}^T = \mathbf{G}$. Since Eq.~(\ref{eq:min J}) represents a linear optimization problem a unique solution exists. The concept behind minimizing the reconstruction error, expressed by \eqref{eq:reconstruction error}, is shown Fig.~\ref{autoencoder}. The left side of the Fig.~\ref{autoencoder} seeks to find a low-dimensional representation of vector $\bm{\phi}_k \in \mathbb{R}^M$, denoted by $\bm{\phi}^s_k \in \mathbb{R}^P$ for $P < M$. The right side projects the low-dimensional representation vector $\bm{\phi}^s_k$ into a high-dimensional space $\tilde{\bm{\phi}}_k = \mathbf{\tilde{G}}^T\bm{\phi}^s_k 
   \in \mathbb{R}^M $. 

The columns of matrix $\mathbf{\tilde{G}}$ are assumed to be orthonormal. This implies:
\begin{eqnarray}\label{eq:orthonormal vec}
&& \mathbf{\tilde{g}}^T_i\mathbf{\tilde{g}}_j = 0, \hspace{0.5cm} i \neq j \\
&& \mathbf{\tilde{g}}^T_i\mathbf{\tilde{g}}_i = 1 
\end{eqnarray}

The orthonormality translates into scaling, which implies that we loose the track of the absolute values of the projected phase noise source vector $\bm{\phi}^s_k$. In short, once the matrix $\mathbf{\tilde{G}}^T = \mathbf{G}$ that solves the minimization problem defined in \eqref{eq:min J} has been found, by applying it directly in \eqref{eq:inverse model compact} will result in scaled versions of the phase noise sources $\bm{\phi}^s_k$. Proper scaling factor will therefore need to be found to obtain true values of the independent phase noise sources. This implies that if we are using the reconstruction error method to find the projection matrix $\mathbf{G}$, \eqref{eq:inverse model compact} needs to be modified accordingly:

\begin{equation}\label{eq:norm_receovery_eq}
  \bm{\phi}^s_k = \mathbf{D} \mathbf{G}\bm{\phi}_k
\end{equation}

where $\mathbf{D}$ is a diagonal matrix containing scaling coefficients which need to be determined.  

It has been shown in \cite{murphy2013machine}, that the solution to the minimization problem defined in Eq.~(\ref{eq:min J}) is obtained by setting $\mathbf{G}= \mathbf{Q}^T_P$, where matrix $\mathbf{Q}_P$ contains $P$ eigenvectors of the sample (square) covariance matrix $\mathbf{S}\in \mathbb{R}^{M\times M}$ obtained from $k=K$ samples:

\begin{eqnarray}\label{eq:corvariance}
   \mathbf{S}(K) &=& \frac{1}{K-1} \sum_{k=1}^K \bm{\phi}_k\bm{\phi}^T_k \nonumber \\
    &=& \mathbf{H}\left(\frac{1}{K-1}\sum_{k=1}^K\bm{\phi}^s_k{\bm{\phi}^s}^T_k\right)\bm{H}^T = \mathbf{H} \mathbf{R}_{\phi}^s\mathbf{H}^T
\end{eqnarray}

where $\mathbf{R}^s_{\phi}$ is a sample covariance matrix of the phase noise sources obtained from $K$ samples. If the phase noise sources are independent that also means they are uncorrelated. This implies that the matrix $\mathbf{R}^s_{\phi}$ is diagonal. The diagonal elements are the variances of the different comblines calculated from $k=0$ to $k=K$. Letting $\sigma^{2}_{i_K}$ represent this variance of the $i$'th combline means that $\mathbf{R}^s_{\phi}$ takes the form:

\begin{equation}\label{eq:R_phi}
    \mathbf{R}^s_{\phi} =
    \begin{bmatrix}
    \sigma^{2}_{1_K} & \hdots & 0\\
    \vdots & \ddots & \vdots \\
    0 & \hdots & \sigma^{2}_{P_K}\\
    \end{bmatrix}
\end{equation}
 
 For non-stationary processes $\sigma^{2}_{i_K}$ has a non-trivial relation to the process variance at $ K=1 $, defined as $\sigma^{2}_{i}$:

\begin{align}
    \sigma^{2}_{i_K} = \frac{1}{K-1}\sum_{k=1}^{K} \alpha_i(k,T_s) \sigma^2_i,
\end{align}

where $ \alpha_i(k,T_s) $ is a factor describing the time evolution of the process' variance. When the phase noise sources are described as Wiener processes then $ \alpha_i(k,T_s)=k $, so the sum over $ k $ in \eqref{eq:corvariance} becomes:

\begin{align}\label{eq:var_Wiener}
    \sigma_{i_K}^2 &= \frac{1}{K-1} \sum_{k=1}^{K} k \, \sigma^2_i \nonumber\\
	&= \sigma^2_i \frac{K}{2}\frac{K+1}{K-1}\nonumber \\
	&\approx \sigma^2_i \frac{K}{2},
\end{align}
where in the last step, we have used that the rightmost fraction is approximately one for large $ K $. Thus, assuming a Wiener process, $ \mathbf{R}^s_{\phi} $ of \eqref{eq:corvariance} can be written as:

\begin{equation}\label{eq:R_phi_Wiener}
    \mathbf{R}^s_{\phi} =
     \frac{K}{2}\begin{bmatrix}
    \sigma^{2}_{1} & \hdots & 0\\
    \vdots & \ddots & \vdots \\
    0 & \hdots & \sigma^{2}_{P}\\
    \end{bmatrix}
\end{equation}

In order to keep things as general as possible, we use covariance, $ \mathbf{R}^s_{\phi} $, matrix formulation given by \eqref{eq:R_phi}. The sample correlation matrix in \eqref{eq:corvariance} is then expressed as:

\begin{equation}\label{eq:Sx}
    \mathbf{S}(K) = \sum_{i=1}^P\sigma^2_{i_K}\mathbf{h}_i\mathbf{h}^T_i
\end{equation}

The matrix $\mathbf{S}(K)$ can then be factored into:

\begin{eqnarray}\label{eq:S_QLQ}
    \mathbf{S} =\mathbf{Q}\bm{\Lambda}\mathbf{Q}^T \nonumber \\
     = [\mathbf{q}_1,...,\mathbf{q}_P,\mathbf{q}_{P+1},...,\mathbf{q}_M]
     \begin{bmatrix}
    \lambda_1 & \hdots & 0\\
    \vdots & \lambda_P & \vdots \\
    0 & \hdots & 0\\
    \end{bmatrix}  \\ \nonumber
   \cdot [\mathbf{q}_1,...,\mathbf{q}_P,\mathbf{q}_{P+1},...,\mathbf{q}_M]^T \nonumber  \\
     = [\mathbf{Q}_P,\mathbf{Q}_{M-P}]\bm{\Lambda}[\mathbf{Q}_P,\mathbf{Q}_{M-P}]^T = \sum_{i=1}^M \lambda_i \mathbf{q}_i\mathbf{q}^T_i = \sum_{i=1}^P \lambda_i \mathbf{q}_i\mathbf{q}^T_i\nonumber
\end{eqnarray}\label{eq:S factored}
    
where $\mathbf{q}_i$, for $i=1,...,M$ and $\lambda_i$, for $i=1,...,P$ are eigenvectors and eigenvalues associated with matrix $\mathbf{S}$, respectively. The rank of matrix $\mathbf{S}$ is $P$ which means that $\mathbf{S}$ has $P$ non-zero and $(M-P)$ zero eigenvalues, i.e.~$ \lambda_i = 0 $ for $i=P+1,...,M$. This implies that the number of non-zero eigenvalues of matrix $\mathbf{S}$ corresponds to the number of phase noise sources in \eqref{eq:elastic tape model gen}. \vspace{0.1cm}

Combining Eq.~(\ref{eq:Sx}) and (\ref{eq:S_QLQ}), we get

\begin{equation}\label{eq:SeQ}
\sum_{i=1}^P\sigma^2_{i_K}\mathbf{h}_i\mathbf{h}^T_i = \sum_{i=1}^P \lambda_i \mathbf{q}_i\mathbf{q}^T_i
\end{equation}

It is observed from \eqref{eq:SeQ} that vectors from generation matrix, $\mathbf{h}_i$ are related to the eigenvectors $\mathbf{q}_i$. Therefore, by plotting the $M$ dimensional eigenvectors $\mathbf{q}_i$, for $i=1,...,P$, as a function of comb line number $m=-(M-1)/2$ to $(M-1)/2$, we will be able to identify how different phase noise sources contribute to the overall phase noise of different comb lines. 

Since $\mathbf{Q}^T\mathbf{Q}=\mathbf{I}$ is orthonormal, this implies that $\parallel \mathbf{q}_i \parallel$=1. We therefore perform the following normalization of \eqref{eq:SeQ}:

\begin{equation}\label{eq:SeQ_2}
    \sum_{i=1}^P\sigma^2_{i_K}\parallel \mathbf{h}_i \parallel^2 \frac{\mathbf{h}_i}{\parallel \mathbf{h}_i\parallel}\frac{\mathbf{h}^T_i}{\parallel \mathbf{h}_i\parallel} = \sum_{i=1}^P \lambda_i \mathbf{q}_i\mathbf{q}^T_i 
\end{equation}

The eigenvalues are then expressed as:

\begin{equation}\label{eq:lambda_i}
    \lambda_i = \sigma^2_{i_K} \parallel \mathbf{h}_i\parallel^2
    \implies \sigma^2_{i_K} = \frac{\lambda_i} {\parallel \mathbf{h}_i\parallel^2} 
\end{equation}

Equation (\ref{eq:lambda_i}) expresses that the variance of the $i^{th}$ independent phase noise source $\sigma^2_{s,i}$ can be obtained from the eigenvalue $\lambda_i$ if the column vector, $\mathbf{h}_i$, of the generation matrix is known. By plotting the normalized eigenvalues $\lambda_i/\parallel \mathbf{h}_i\parallel^2$ as a function of time $KT_s$ for increasing $K$, a time-evolution of the variance of the independent phase noise sources is obtained. In that way, we can get variance of each individual phase noise source. The total variance is obtained by summing up all eigenvalues. 

The phase noise source described as a Wiener process will have a Lorentzian spectrum of Full Width Half Maximum (FWHM) $\Delta \nu_i$. Its phase variance can then be expressed at time $T = KT_s$ according to:

\begin{equation}\label{eq:var_PN}
    \sigma^2_{i_K} = \frac{2 \pi \Delta \nu_i T}{2} 
\end{equation}

The division by two comes from the defining $ \sigma^2_{i_K}  $ as the sample covariance of the phase instead of the variance of their steps $ \sigma_i $. This difference is found in \eqref{eq:var_Wiener}. For a Wiener process, we can thereby directly relate spectral broadening (linewidth) of $i^{th}$ independent phase noise source to its eigenvalue:

\begin{equation}\label{eq:lambda2LW}
    \Delta \nu_i = \frac{\lambda_i  }{\pi T \mathbf\parallel \mathbf{h}_i\parallel^2}
\end{equation}

In order to recover phase noise sources, $\bm{\phi}^s_k$, we apply a matrix $\mathbf{Q}_P$ containing eigenvectors associated with $P$ independent phase noise sources, $\mathbf{Q}_P=[\mathbf{q}_1,...,\mathbf{q}_P]$, in \eqref{eq:norm_receovery_eq}: 

\begin{equation}\label{eq:norm_receovery_eq_Q_P}
  \bm{\phi}^s_k = \mathbf{D} \mathbf{Q}^T_P\bm{\phi}_k
\end{equation}

Since the eigenvectors, $\mathbf{q}_i$, are orthonormal,  the diagonal denormalization matrix $\mathbf{D}$, is found by performing the following normalization of \eqref{eq:SeQ}

\begin{equation}\label{eq:SeQ_normQ}
\sum_{i=1}^P\sigma^2_{i_K}\mathbf{h}_i\mathbf{h}^T_i = \sum_{i=1}^P \frac{\lambda_i}{\parallel \mathbf{h}_i \parallel^2} \mathbf{q}_i\parallel \mathbf{h}_i \parallel\mathbf{q}^T\parallel \mathbf{h}_i \parallel
\end{equation}

This implies that the denormalization matrix is expressed as:

\begin{equation} \label{eq:normalization_matrix_D}
\mathbf{D} =
 \begin{bmatrix}
    \parallel \mathbf{h}_1 \parallel& \hdots & 0\\
    \vdots & \ddots & \vdots \\
    0 & \hdots & \parallel \mathbf{h}_P \parallel\\
    \end{bmatrix}
\end{equation}




\begin{figure*}[t!]
\centering \includegraphics[width=\textwidth]{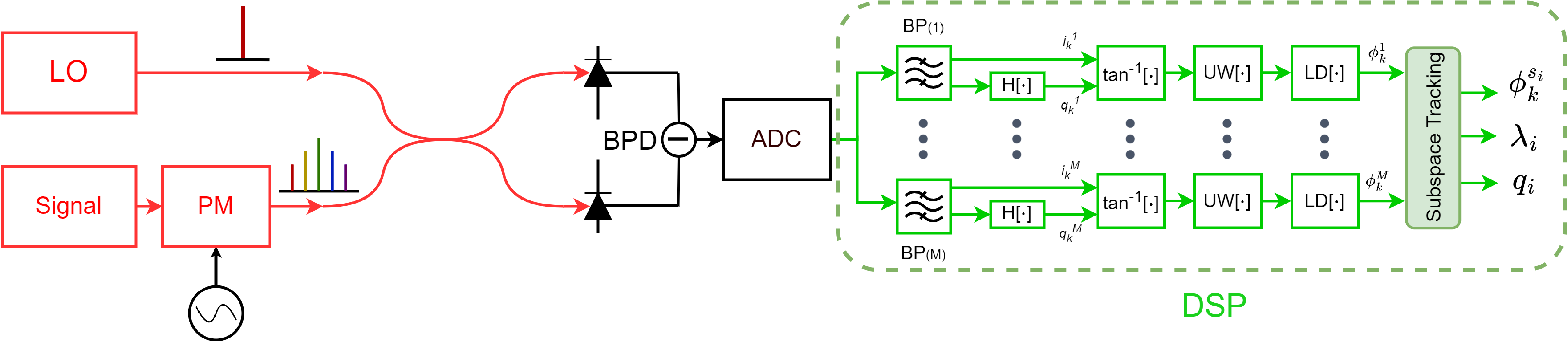}
\caption{Employed set-up for numerical and experimental investigations of the proposed sub-space tracking method to identify independent sources of phase noise associated with an EO comb. BPF: bandpass filter, H$[\cdot]:$ Hilbert transform, UW: unwrapping, LD: linear detrending.}
\label{setupeo}
\end{figure*}

So far, we have assumed that the noise term in Eq.~(\ref{eq:signal_coh_det}) was negligible, i.e.~$n_k\approx 0$. However, this is not the case in practice. Indeed, the noise term $n_k$ will results in a noise floor which sets a limit for the phase noise estimation. If the phase noise sources have magnitudes below the noise floor, they will not be detectable. If optimal filtering is employed, the noise term, $n_k$ can be minimized as shown in \cite{Zibar2021}. However, the noise term, $n_k$, can never be fully eliminated, and the estimated phase noise will thereby always contain residual measurement noise $n^r_k$ \cite{Zibar2021}. This implies that \eqref{eq:forward model compact} needs to be modified to:

\begin{equation}\label{eq:linear_model_noise}
    \bm{\phi}_k = \mathbf{H}\bm{\phi}^s_k + n^r_k\mathbf{I} = \sum_{i=1}^P\mathbf{h}_p\phi^{s_i}_k + n^r_k\mathbf{I}
\end{equation}

where $\mathbf{I}$ is a $M\times M$ identity matrix and the residual measurement noise is sampled from Gaussian distribution, i.e.~ $n^r_k \sim N(0,\sigma_r^2)$ and $\sigma^2_r$ is the variance of the residual measurement noise. In the case of shot-noise limited balanced detection and Kalman filtering based optical phase estimation, the measurement noise will take its minimum value \cite{Zibar2021}.   

In the presence of residual measurement noise, the covariance matrix in Eq.~(\ref{eq:corvariance}) then takes the following form:

\begin{equation}\label{eq:corvariance_noise}
    \mathbf{S}(K) =  \mathbf{H} \mathbf{R}^s_{\phi}\mathbf{H}^T
  +   \sigma^2_r \mathbf{I}
\end{equation}

The correlation matrix of the additive white Gaussian noise process $\mathbf{R}_n = \sigma^2_r\mathbf{I}$ has a single degenerate eigenvalue $\lambda = \sigma^2_r$ with multiplicity $M$. This implies that any vector qualifies as eigenvector. The eigenvalue decomposition can now be expressed as:

\begin{eqnarray}\label{eq:S_with_noise}
 \mathbf{S}(K) =   \\ \nonumber
 [\mathbf{Q}_P,\mathbf{Q}_{M-P}] \\ \nonumber  
   \cdot \left(    \begin{bmatrix}
    \lambda_1 & \hdots & 0\\
    \vdots & \lambda_P & \vdots \\
    0 & \hdots & 0\\
    \end{bmatrix}
    +
       \begin{bmatrix}
    \sigma^2_r & \hdots & 0\\
    \vdots & \ddots & \vdots \\
    0 & \hdots & \sigma^2_r\\
    \end{bmatrix}\right)  \\ \nonumber
  \cdot  [\mathbf{Q}_P,\mathbf{Q}_{M-P}]^T = \mathbf{Q}\mathbf{\tilde{\Lambda}}\mathbf{Q}^T
\end{eqnarray}

It should be noticed that the residual noise power, $\sigma^2_r$, is uniformly distributed in the entire Euclidean space, while the original phase noise contributions are constrained to $P$ dimensions. The last $(M-P)$ eigenvalues of the diagonal matrix $\mathbf{\tilde{\Lambda}}$ contain information about the variance of the residual measurement noise. If the residual measurement noise is additive white Gaussian noise, then  the last $(M-P)$ eigenvalues should be independent of the discrete time $KT_s$. By computing $M$ eigenvalues of the diagonal matrix $\mathbf{\tilde{\Lambda}}$, we will be able to determine if the eigenvalues are dominated by the variance of the independent phase noise source or the residual measurement noise. Since, phase noise is a non-stationary process, its variance will increase with time and will dominate over the residual measurement noise, i.e. $\sigma^2_{i_K} \gg \sigma^2_r$. This implies that we can still use eigenvalues in \eqref{eq:lambda_i} to evaluate the variance of the independent phase noise sources. 

Finally, even in the presence of the residual measurement noise, eigenvectors $\mathbf{q}_i$, for $i=1,...,P$, associated with sample covariance matrix  \eqref{eq:S_with_noise} can still be used to estimate the independent phase noise sources by applying \eqref{eq:norm_receovery_eq_Q_P}. However, as it will be shown using numerical simulations and experiments, the measurement noise will eventually set a limitation on the dynamic range for the phase noise estimation. This will be especially pronounced if the phase noise sources have very low magnitude as the measurement noise floor will be approached fast. \vspace{0.2cm}






Performing subspace tracking relies on accurate estimation of the sample covariance matrix $\mathbf{S}(k)$, for $k=1,...,K$. When operating in a nonstationary environment, which is the case of phase noise, exponential windowing needs to be employed \cite{Adali2010}:

\begin{equation}\label{eq:S_exp_win}
    \mathbf{S}(K) = \frac{1}{K-1}\sum_{k=1}^K \beta^{K-k}\bm{\phi}_k\bm{\phi}^T_k
\end{equation}

where $0<\beta<1$ is the forgetting factor and it ensures that the samples in the distant past are downweighted, which is important for the nonstationary signals. The sample covariance matrix can be updated recursively as:

\begin{equation}\label{eq:S_w_iter}
    \mathbf{S}(k) = \beta\mathbf{S}(k-1) +\bm{\phi}_k\bm{\phi}^T_k
\end{equation}

where $k=1,...,K$. In general, iterative computation of the sample covariance matrix is recommended due to computational efficiency. 

\section{Numerical results}
\label{sec:num_results} 

In this section, the theoretical framework presented in Section \ref{sec:theory} is applied to perform phase noise source identification and separation from an electro-optic comb. Electro-optic combs are well understood and simple to model and can therefore be used to verify the proposed framework. \vspace{0.1cm}


Detailed numerical investigations based on the set-up depicted in Fig.~\ref{setupeo} are performed. The entire simulation environment is implemented on the component level to resemble the experiment that is discussed in section \ref{sec:exp_results}.   

As shown in Fig.~\ref{setupeo}, a multi-heterodyne coherent set-up employing two optical sources (CW signal and the CW LO), RF signal source, phase modulator (PM), 3 dB coupler, balanced photodetector (BPD) and ADC is used. The phase noise of the CW signal, $\phi^{sig}_k$, and LO laser, $\phi^{LO}_k$, as well as the RF signal source, $\phi^{RF}_k$, is modelled as a Wiener process with the corresponding linewidths of: $\Delta\nu_{sig} =\Delta\nu_{LO}= 10$ kHz and $\Delta\nu_{RF} = 1$ kHz, respectively. 

The flicker or random walk noise is not simulated as it would be embedded into the total phase noise of the CW laser and RF oscillator. Additionally, for real-world RF oscillators, the phase noise of RF oscillators is far away from a Wiener process. However, we would like to point out that the framework does not set any restrictions on the distribution of phase noise sources. The only requirement for the method to be able to separate the phase noise sources is that they obey Eq.~(\ref{eq:elastic tape model gen}).

The frequency and the power of the RF source is 500 MHz and 32 dBm, respectively. The $V_{\pi}$ of the phase modulator is 3 V. The power of the signal and the LO laser is set to 20 dBm, and the sampling frequency of the ADC is $F_s=40$ GHz. The number of samples used for the simulation is $K=40\times10^6$\vspace{0.1cm}




First, the generated EO comb and the LO laser are combined in a 3 dB coupler prior to balanced photodetection and sampling. The detected and sampled beat signal is a downconverted comb, as expressed by Eq.~(\ref{eq:signal_coh_det}), with the  frequency spacing, $\Delta f=500$ MHz and $f_0=0$.



As indicated in Fig.~\ref{setupeo}, we first employ phase estimation  prior to subspace tracking and the subsequent estimation of the phase noise sources.
The phase estimation method involves using a bank of brick-wall bandpass (BP) filters, with bandwidth of $B=500$ MHz, to extract different frequency comb lines from the detected beat signal. For each filtered line $i_k^m$, a Hilbert transform is applied to recover the corresponding orthogonal quadrature $q_k^m$. The phase of the $mth$ comb-line is then estimated by taking the inverse tangent, i.e. $\tan^{-1}(i_k^m/q_k^m)$, of the reconstructed field. The resulting phases are processed through an unwrapping function that smooths out any large jumps in phase greater than $\pi$ by adding multiples of $2\pi$. This produces a continuous function of the phase over time. Finally, a linear detrending is applied to the phase traces to remove frequency dependence and produce estimates of the phase noise $\phi_k^m$ for each of the $M$ comb lines. Obtained phase noise traces $\phi^m_k$ can be then used to construct the sample covariance matrix, compute eigenvectors, and eigenvalues, and finally estimate the sources of phase noise associated with an EO comb. \vspace{0.1cm}  

According to \cite{Ishizawa:13}, the phase noise of an electro-optic comb lines can be described by the first two terms in \eqref{eq:elastic tape model gen}). This implies that the phase noise of the detected comb, using the set-up in Fig .~\ref{setupeo} can be expressed as:

\begin{equation} \label{eq:beat_eo_comb_sim}
    \phi^m_k = \phi^{sig}_k-\phi^{LO}_k - m\phi^{RF}_k
\end{equation}

where $m=-(M-1)/2,...,(M-1)/2$. On matrix form, \eqref{eq:beat_eo_comb_sim} is expressed as:

\begin{equation}\label{eq:forward model matrix EO}
    \begin{bmatrix}
    \phi^{-(M-1)/2}_k \\
    \vdots \\
    \phi^{(M-1)/2}_k
    \end{bmatrix}=
     \overbrace{\begin{bmatrix}
    1 & -1 & -(M-1)/2\\
    \vdots & \vdots & \vdots \\
    1 & -1 & (M-1)/2\\
    \end{bmatrix}}^{\mathbf{H}}
    \begin{bmatrix}
    \phi^{sig}_k \\
     \phi^{LO}_k \\
    \phi^{RF}_k
    \end{bmatrix}
\end{equation}

According to \eqref{eq:forward model matrix EO}, the phase noise contribution originating from the CW signal and LO source, $\phi^{sig}_k$ and $\phi^{LO}_k$, will transfer equally to all comb-lines. The phase noise contribution from the RF source, $\phi^{RF}_k$, will scale linearly with comb-line number $m$.  

The first two column vectors of matrix $\mathbf{H}$, in \eqref{eq:forward model matrix EO}, ($\mathbf{h}_1$ and $\mathbf{h}_2$), are linearly dependent, while the third column vector $\mathbf{h}_3$ is linearly independent with respect to  $\mathbf{h}_1$ and $\mathbf{h}_2$. The rank of $\mathbf{H}$ in Eq.~(\ref{eq:forward model matrix EO}) is therefore 2. This means that we can only identify two phase noise sources. Since, we only have two sources of phase noise, we choose to process $M=5$ comb lines, $\bm{\phi}_k=[\phi^{-2}_k,..\phi^{0}_k...,\phi^2_k]^T$. In general, the number of detected comb-lines $M$ needs to be chosen such that it is greater than the expected number of independent phase noise sources.   

Rewriting, Eq.~(\ref{eq:forward model matrix EO}), in a reduced form with full rank of 2, we obtain: 

\begin{equation}\label{eq:forward model matrix EO 2}
    \begin{bmatrix}
    \phi^{-(M-1)/2}_k \\
    \vdots \\
    \phi^{(M-1)/2}_k
    \end{bmatrix}=
     \overbrace{\begin{bmatrix}
    1 &  -(M-1)/2\\
    \vdots &  \vdots \\
    1 &  (M-1)/2\\
    \end{bmatrix}}^{\mathbf{H}}
    \begin{bmatrix}
    (\phi^{sig}_k-\phi^{LO}_k) \\
    \phi^{RF}_k
    \end{bmatrix}
\end{equation}

Assuming that \eqref{eq:forward model matrix EO} is an accurate model to describe phase noise of an EO comb, We can now use, Eq.~(\ref{eq:matrixG_first_order}) to obtain the projection matrix $\mathbf{G}$, and recover the phase noise sources, i.e.~$(\phi^{sig}_k-\phi^{LO}_k)$ and $\phi^{RF}_k$. The first recovered phase noise source, $(\phi^{sig}_k-\phi^{LO}_k)$, is associated with the phase noise of the beat between the CW laser signal and the LO. The second recovered independent phase noise source is associated with the RF signal generator, i.e.~$\phi^{RF}_k$. In order to characterize EO comb phase noise sources, ($\phi^{sig}_k$ and $\phi^{RF}_k$), we either need to use LO that has significantly lower phase noise than the signal laser or have signal and LO lasers with similar phase noise performances. For the numerical simulations, we assume that the signal and LO laser have the same linewidths, and $\phi^{sig}_k$ is obtained by dividing the estimated beat phase noise, $(\phi^{sig}_k - \phi^{LO}_k)$, by factor 2. \vspace{0.1cm}

\begin{figure}[h!]
\centering \includegraphics[width=0.7\linewidth]{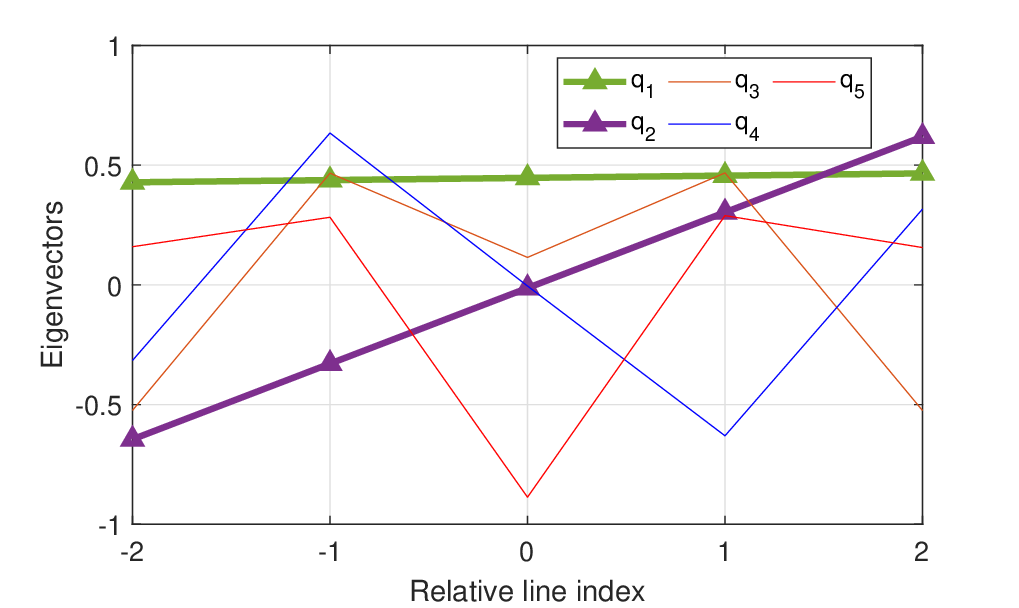}
\caption{(No measurement noise) Evolution of eigenvectors, of sample correlation matrix $\mathbf{S}(K)$, as a function of relative line index $m$ for $K=40\times 10^6$.}
\label{fig:eigVec_vs_index_sim_noNoise}
\end{figure}

A more general approach to separate and estimate the phase noise sources that does not rely on any prior knowledge about phase noise generation process (generation matrix agnostic), is achieved by performing eigenvalue decomposition of the correlation matrix $\mathbf{S}(K)$ and then employing \eqref{eq:norm_receovery_eq_Q_P}. 

By plotting the eigenvectors $[\mathbf{q}_1,...,\mathbf{q}_M]$, we will be able to identify the number of independent phase noise sources and their scaling with comb-line number. This information is then used to construct the normalization matrix $\mathbf{D}$ in \eqref{eq:normalization_matrix_D}. \vspace{0.2cm}

\begin{figure}[h!]
\centering \includegraphics[width=0.7\linewidth]{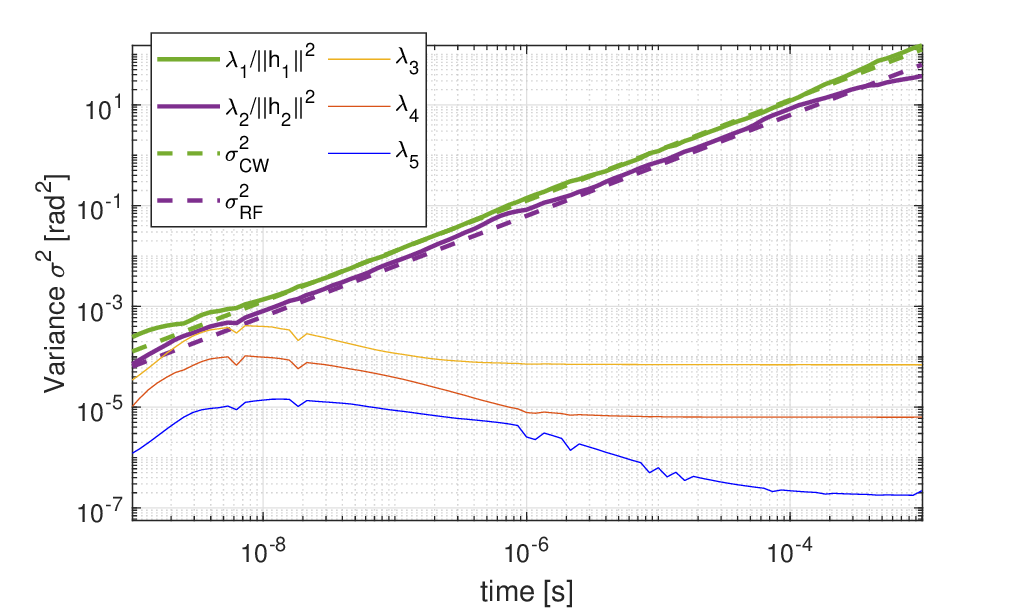}
\caption{(No measurement noise) Evolution of scaled eigenvalues (variance), of sample correlation matrix $\mathbf{S}(k)$, for $k=1,...,40\times 10^6 $, as a function of time $t = KT_s$.}
\label{fig:lambda_vs_t_sim}
\end{figure}

In Fig.~\ref{fig:eigVec_vs_index_sim_noNoise}, the evolution of $M=5$ eigenvectors $[\mathbf{q}_1,...,\mathbf{q}_5]$ is plotted as a function of comb-line number $m$ for $K=40 \times 10^6$. It is observed that the first eigenvector, $\mathbf{q}_1$,  is constant as a function of comb-line number $m$, while the second eigenvector $\mathbf{q}_2$ increases linearly. The remaining three eigenvectors, $[\mathbf{q}_3,...,\mathbf{q}_5]$ show no structure and exhibit more or less random fluctuations, which can be attributed to the numerical inaccuracies (noise) when performing eigenvalue decomposition. Since, eigenvectors are related to the generation matrix by \eqref{eq:SeQ_normQ}, the evolution of the eigenvectors in Fig.~\ref{fig:eigVec_vs_index_sim_noNoise}, confirms that the phase noise generation matrix $\mathbf{H}$ in \eqref{eq:forward model matrix EO 2} is accurate for describing phase noise generation process of the EO comb under the consideration.   \vspace{0.2cm}

Next, we plot scaled eigenvalues, $[\lambda_1,...,\lambda_5]$, associated with the sample covariance matrix $\mathbf{S}(k)$, for $k=1,...K$, as a function of time $KT_s$ in Fig.~\ref{fig:lambda_vs_t_sim}. The scaling of $\lambda_1$ and $\lambda_2$ is performed in order to convert eigenvalues to variances according to \eqref{eq:lambda_i}.

The first observation to be made is that the first two scaled eigenvalues (variances), $\sigma^2_{1_K}=\lambda_1/\parallel \mathbf{h}_1 \parallel^2$ and $\sigma^2_{2_K}=\lambda_1/\parallel \mathbf{h}_2 \parallel^2$, associated with the first two eigenvectors, $\mathbf{q}_1$ and $\mathbf{q}_2$ increase linearly with the time, while the remaining eigenvalues remain more or less constant. This is very much inline with Fig.~\ref{fig:eigVec_vs_index_sim_noNoise}, and \eqref{eq:forward model matrix EO 2}, indicating that the EO comb is dominated by only two sources of phase noise. Finally, we observe that an overlap is obtained between $\sigma^2_{1_K}$ and $\sigma^2_{2_K}$, and variances of the phase noise of the CW laser signal and RF source i.e.~$\sigma^2_{sig}$ and $\sigma^2_{RF}$ This is fully in accordance with \eqref{eq:lambda_i}.     

Ideally, the eigenvalues $[\lambda_3,..,\lambda_5]$ should theoretically be zero. However, due to the numerical imprecision of the algorithm used to perform eigenvalue decomposition, they are not. Since  $[\lambda_3,..,\lambda_5]$ do not have a physical meaning, we have scaled them with factor 1 to convert them to variances.  \vspace{0.2cm}

Finally, we use \eqref{eq:norm_receovery_eq_Q_P} and \eqref{eq:normalization_matrix_D} to estimate the phase noise sources associated with the EO comb, by employing the following matrix multiplication: 

\begin{equation}\label{eq:reverse model matrix EO 2}
    \begin{bmatrix}
    \phi^{s_1}_k \\
    \phi^{s_2}_k
    \end{bmatrix}=
     \begin{bmatrix}
    \mathbf{q}_1\cdot||\mathbf{h}_1||, \mathbf{q}_2\cdot||\mathbf{h}_2||\\
    \end{bmatrix}
    \begin{bmatrix}
    \phi^{-2}_k \\
      \vdots \\
    \phi^{2}_k
    \end{bmatrix}
\end{equation}

\begin{figure}[h!]
  \centering
    \begin{subfigure}{0.49\linewidth}
        \includegraphics[width=\linewidth]{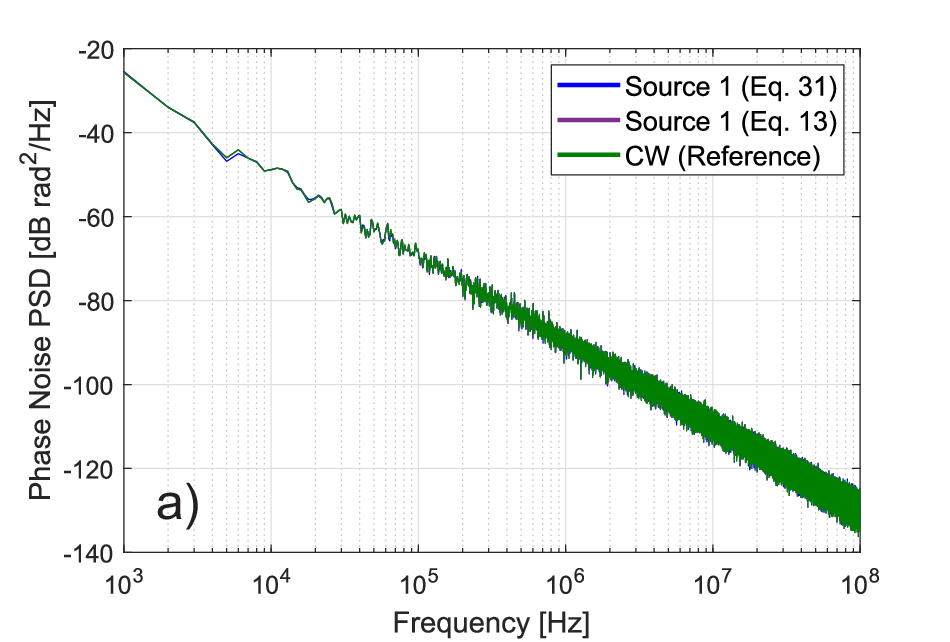}
    \end{subfigure}
    \begin{subfigure}{0.49\linewidth}
    \includegraphics[width=\linewidth]{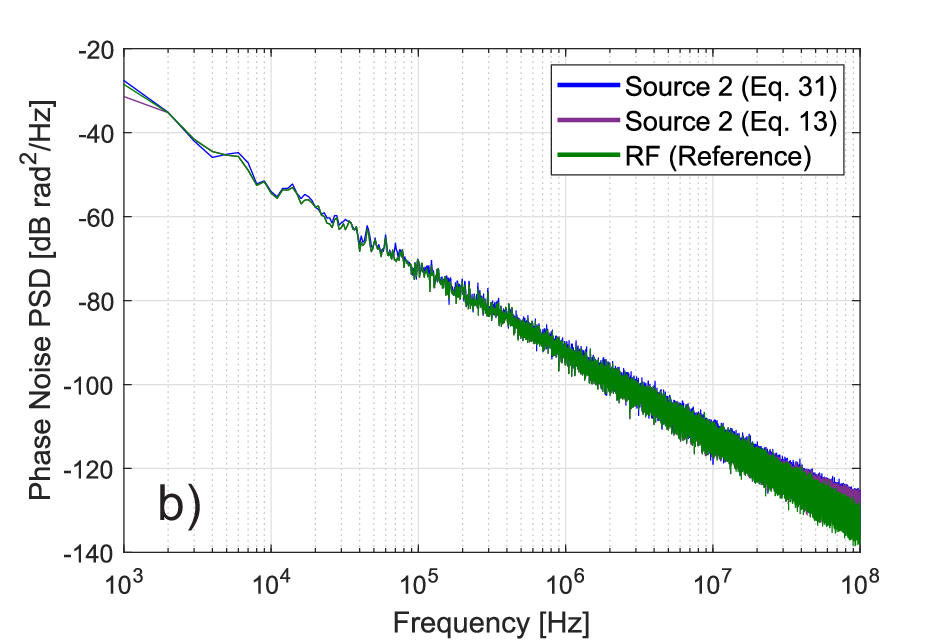}
    \end{subfigure}	
\caption{(No measurement noise) Power spectral density as a function of frequency for (a) the recovered first phase noise source and signal CW laser (b) the recovered second phase noise source and RF signal generator}
\label{fig:PSD_EO_SIMULATION_NO_NOISE}
\end{figure}

In Fig.~\ref{fig:PSD_EO_SIMULATION_NO_NOISE}(a) and (b), the power spectral density (PSD) of the recovered phase noise sources $\phi^{s_1}_k$ and $\phi^{s_2}_k$ is shown. Source 1 and Source 2 indicate phase noise PSD of the estimated phase noise sources $\phi^{s_1}_k$ and $\phi^{s_2}_k$. We also plot phase noise PSD of the CW laser and RF signal generator used for EO comb generation as a reference.   

For comparison reasons, we also use \eqref{eq:matrixG_first_order}, to recover the phase noise sources. It is observed that the recovered phase noise sources Source 1, ($\phi^{s_1}_k$),  and Source 2, ($\phi^{s_2}_k$), fully overlap with the phase noise PSD of the signal CW and RF oscillator, respectively. Once again this is in full agreement with the theoretical phase noise model for EO combs, i.e.~phase noise of the CW laser and the RF oscillator maps directly to the common mode and the repetition rate phase noise of an EO comb. 

For the chosen sampling frequency, $F_s$ and the number of samples, $k=1,..,K$, the resolution frequency, $F_s/K$, (the lowest frequency of the PSD), is 1 kHz. To address, the performance at low frequencies ( $<$ 100 Hz), a solution would be to decrease the sampling rate or increase the number of samples. However, nothing fundamentally changes by decreasing the resolution frequency. We therefore believe that the the framework would also be able to successfully estimate phase noise PSDs in the low-frequency range, as well.

\vspace{0.2cm}

\begin{figure}[h!]
\centering \includegraphics[width=0.5\linewidth]{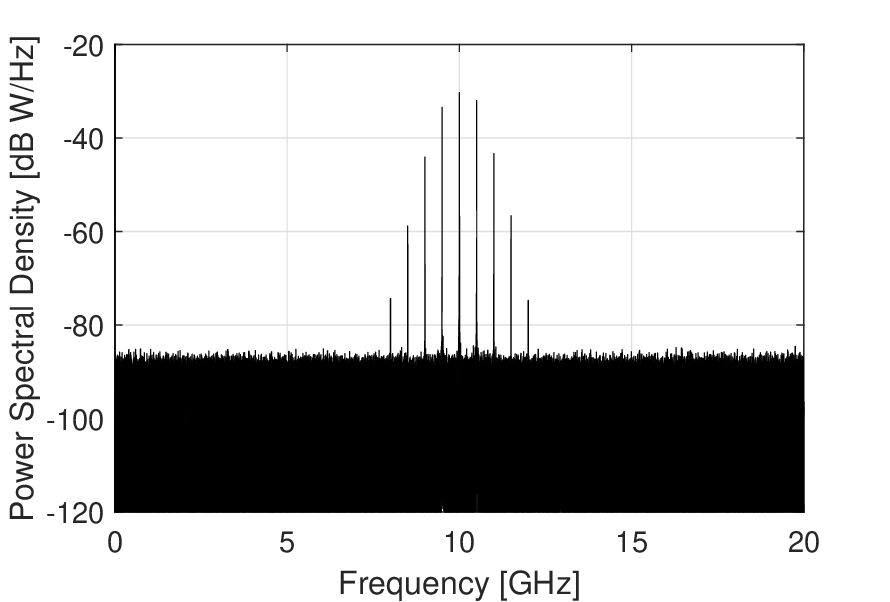}
\caption{(Simulation with measurement noise) Spectrum of the downconverted EO comb.}
\label{fig:spectrum_sim}
\end{figure}

Next, we added noise associated with the receiver front-end depicted in Fig. ~\ref{setupeo}. The noise added in the simulations is white Gaussian noise. The spectrum of the downconverted comb is shown in Fig.~\ref{fig:spectrum_sim}. In Fig.~\ref{fig:spectrum_sim}, only 9 lines are visible. The limited visibility of additional lines can be attributed to their concealment beneath the noise floor. It is worth noting that a high noise floor was employed (surpassing that of real-world scenarios with analogous configurations) to investigate the impact of measurement noise on the framework's outcomes.

\begin{figure}[h!]
\centering \includegraphics[width=0.7\linewidth]{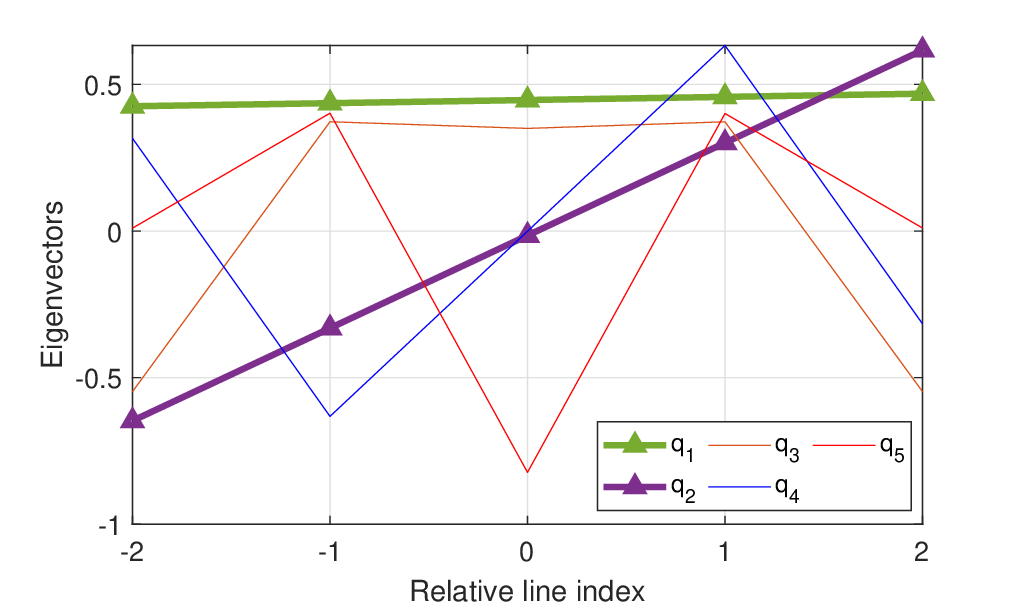}
\caption{(With measurement noise) Evolution of eigenvectors, of sample correlation matrix $\mathbf{S}(K)$, as a function of relative line index $m$ for $K=40\times 10^6$.}
\label{fig:eigVec_vs_index_sim_Noise}
\end{figure}

\begin{figure}[h!]
\centering \includegraphics[width=0.7\linewidth]{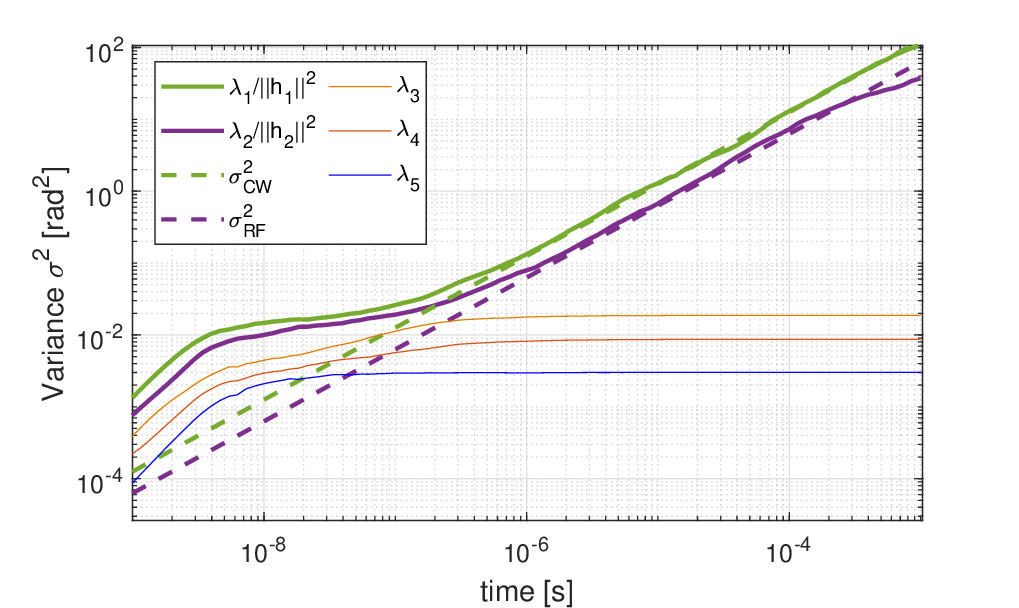}
\caption{(With measurement noise) Evolution of eigenvalues, of sample correlation matrix $\mathbf{S}(k)$, for $k=1,...,40\times 10^6 $, as a function of time $t = KT_s$.}
\label{fig:lambda_vs_t_sim_Noise}
\end{figure}

\begin{figure}[h!]
  \centering
    \begin{subfigure}{0.49\linewidth}
        \includegraphics[width=\linewidth]{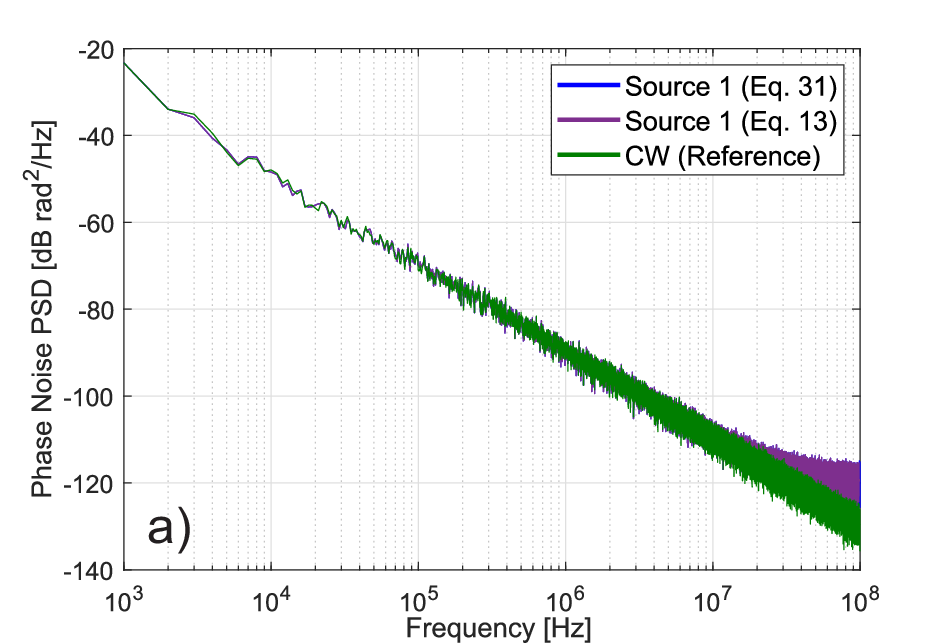}
    \end{subfigure}
    \begin{subfigure}{0.49\linewidth}
    \includegraphics[width=\linewidth]{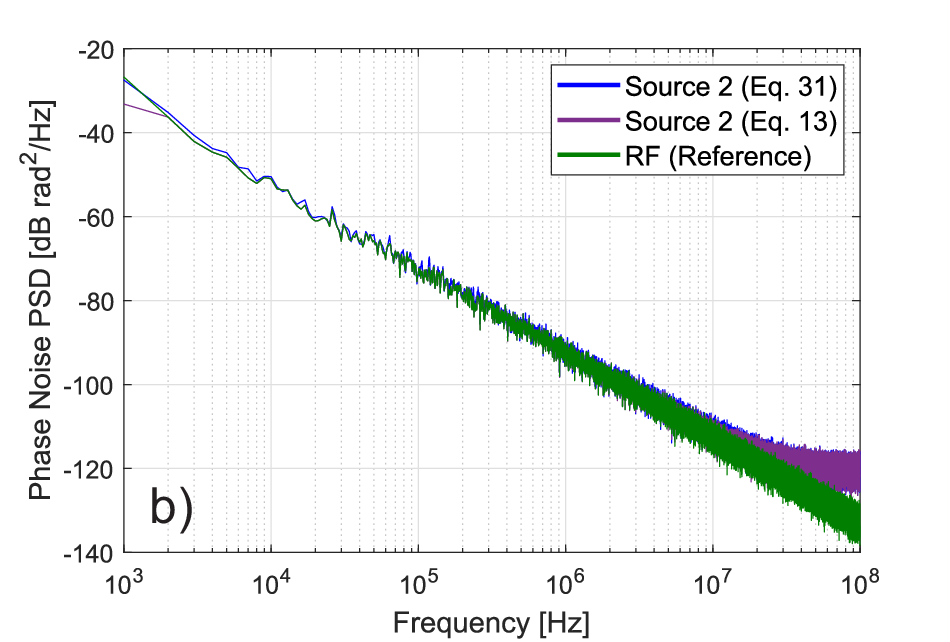}
    \end{subfigure}	
\caption{(With measurement noise) Power spectral density as a function of frequency for (a) the recovered first independent phase noise source and signal CW laser (b) the recovered second independent phase noise source and RF signal generator}
\label{fig:PSD_EO_SIMULATION_NOISE}
\end{figure}

We then repeat the same procedure that we did in the absence of measurement noise, and plot the evolution of eigenvectors, eigenvalues and PSDs, i.e. Fig.~\ref{fig:eigVec_vs_index_sim_Noise}, \ref{fig:lambda_vs_t_sim_Noise} and \ref{fig:PSD_EO_SIMULATION_NOISE}. 
What changes in the presence of measurement noise is that, according to \eqref{eq:S_with_noise}, the last three eigenvalues of the EO comb under consideration, will not be zero but equal to the variance of the measurement noise, $[\lambda_3,...,\lambda_5]=\sigma^2_r$. \vspace{0.1cm}

It is observed that the evolution of eigenvectors in the presence of measurement noise, Fig.~\ref{fig:eigVec_vs_index_sim_Noise}, is the same as in the absence of the measurement noise. Basically, we have two independent phase noise sources, one that is common to all comb-lines and the second one that scales linearly with comb-line number. \vspace{0.1cm}

The evolution of the scaled eigenvalues, representing phase noise variances, in Fig.~\ref{fig:eigVec_vs_index_sim_Noise}, is different compared to Fig.~\ref{fig:eigVec_vs_index_sim_noNoise}. However, overall similar trends can be observed. The reason for discrepancy is that in the presence of measurement noise, variance of the measurement noise is added to the eigenvalues as shown in \eqref{eq:S_with_noise}. This implies that we have $\lambda_1 + \sigma^2_r$ and $\lambda_2 + \sigma^2_r$. Fig.~\ref{fig:lambda_vs_t_sim_Noise} shows that for time up to $10^{-6}$ s, the variance of the measurement noise dominates: $\sigma^2_r >> \{\lambda_1, \lambda_2\}$. That is the reason why we cannot observe the linear evolution of the scaled eigenvalues, $\lambda_1/\parallel \mathbf{h}_1\parallel^2$, and $\lambda_2/\parallel \mathbf{h}_2\parallel^2$ as a function of time. Beyond $10^{-6}$ s, the variance of the independent phase noise sources dominates over the measurement noise.  

\vspace{0.1cm}   
Finally, for the recovered PSD of the independent phase noise sources in Fig.~\ref{fig:PSD_EO_SIMULATION_NOISE}, we see that the measurement noise imposes a noise floor for frequencies above 10 MHz.

\section{Experimental results} \label{sec:exp_results} 
\subsection{Electro-optical frequency comb}

The employed experimental setup for phase noise characterization of the electro-optical frequency comb matches the simulation setup of Fig.~\ref{setupeo}. The setup employs two lasers with similar phase noise profiles, one as a CW signal and the other as the CW LO source. Both optical sources are NKT Photonics Koheras BASIK fiber lasers, with optical output power of 12 dBm. The frequency difference between the two lasers is set to 3 GHz. \vspace{0.2cm}

\begin{figure}[h!]
\centering \includegraphics[width=0.5\linewidth]{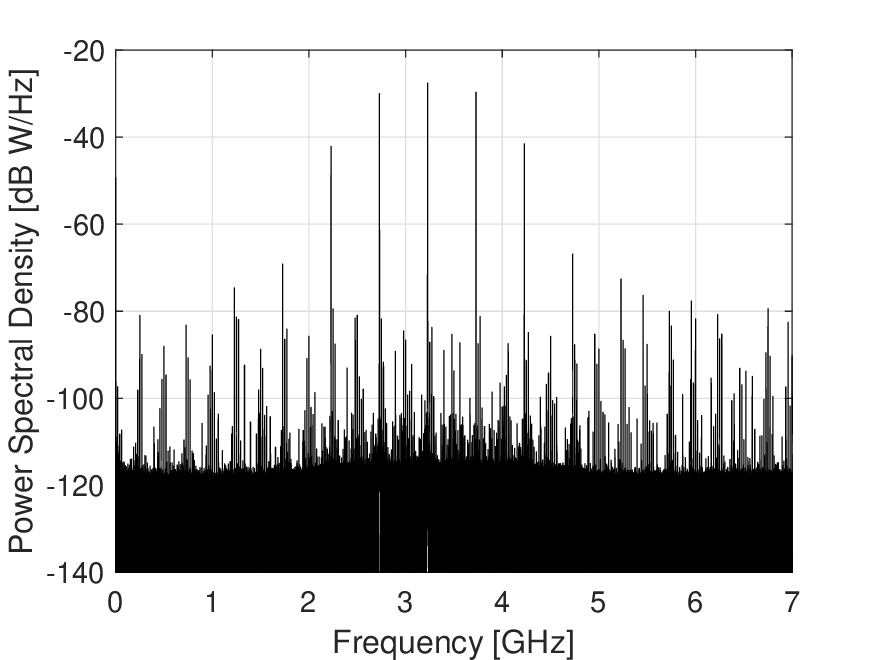}
\caption{(Experimental EO comb) Electrical spectrum of the downconverted EO comb.}
\label{fig:spectrum_exp}
\end{figure}

The output of the CW signal laser is passed though an electro-optical phase modulator, which is driven by the RF signal generator with the frequency $f_{RF} = 500$ MHz. The output power of the RF source is 13 dBm. An electrical low-pass filter (LP), with a bandwidth of 750 MHz is placed after the RF signal source to suppress higher order harmonics. The generated EO comb is combined with an LO laser source prior to balanced detection. The balanced detector has a 3 dB bandwidth of 70 GHz and its output is sampled using a real-time sampling scope with a 3 dB bandwidth of 13 GHz and sampling rate $F_s=40$ GSa/s. The storage capabilities of the scope allow us to save $K = 40\times 10^{6}$ samples. This translates into frequency resolution of $f_{res} = F_{s}/K = 1$ kHz. We employ the same DSP processing stage as for the simulations, see Fig.~\ref{setupeo}. The electrical spectrum of the downconverted comb is shown in Fig.~\ref{fig:spectrum_exp}. The number of the detected comb-lines used for processing is $M=5$. We believe that frequency components below -80 dB W/Hz originate from the spurious tones of the analogue-to-digital converter. \vspace{0.2cm}

Phase noise source separation is achieved by performing eigenvalue decomposition of the correlation matrix $\mathbf{S}(K)$ and then employing \eqref{eq:norm_receovery_eq_Q_P}. The approach is the same as for the numerical simulations and consists of the following steps: 1) BP filtering and phase noise estimation, 2) correlation matrix computation, 3) eigenvalue decomposition and 4) phase noise source separation and estimation, see Fig.~\ref{setupeo}. Since the approach based on the eigenvalue decomposition does not require any prior knowledge about phase noise generation process, it can be used to confirm the theoretical prediction presented in \cite{Ishizawa:13}, experimentally. More specifically, the obtained eigenvectors can be used to infer the generation matrix, $\mathbf{H}$, for that particular frequency comb generation configuration, directly from the experimental data.\vspace{0.2cm}

\begin{figure}[h!]
\centering \includegraphics[width=0.7\linewidth]{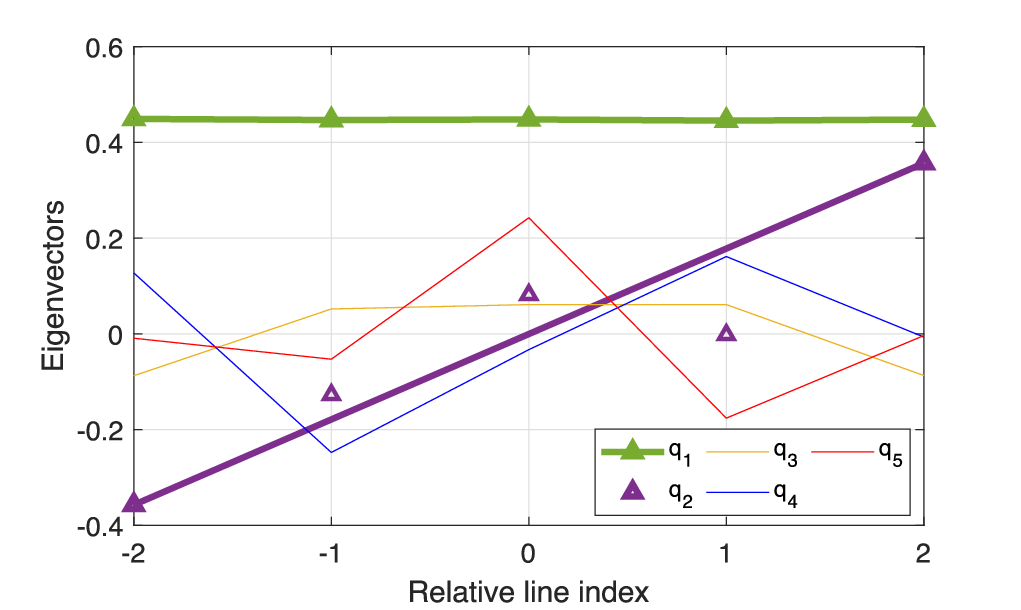}
\caption{(Experimental EO comb) Evolution of eigenvectors, of sample correlation matrix $\mathbf{S}(K)$, as a function of relative line index $m$ for $K=40\times 10^6$.}
\label{fig:eigvec_vs_index_EO_exp}
\end{figure}

In Fig.~\ref{fig:eigvec_vs_index_EO_exp}, the eigenvectors $[\mathbf{q}_1,...,\mathbf{q}_5]$, of the correlation matrix, $\mathbf{S}(K)$, are plotted as a function of comb-line number. The line corresponds to the linear fit of the eigenvector, $\mathbf{q}_2$.
It is observed from Fig.~\ref{fig:eigvec_vs_index_EO_exp}, that for the EO comb under consideration, there are two phase noise sources, represented by the first two eigenvectors. The first eigenvector, $\mathbf{q}_1$, representing the first phase noise source is constant with the comb-line number. The second eigenvector, $\mathbf{q}_2$, representing the second  phase noise source scales linearly with comb-number. This is well in accordance with the numerical investigations and theoretical predictions. This implies that we can clearly identify column of the generation matrix $\mathbf{H}$, i.e.~$\mathbf{h}_1$ and $\mathbf{h}_2$. The remaining eigenvectors $[\mathbf{q}_3,...,\mathbf{q}_5]$ show random fluctuations and do not show any structure. \vspace{0.2cm}  

\begin{figure}[h!]
\centering \includegraphics[width=0.7\linewidth]{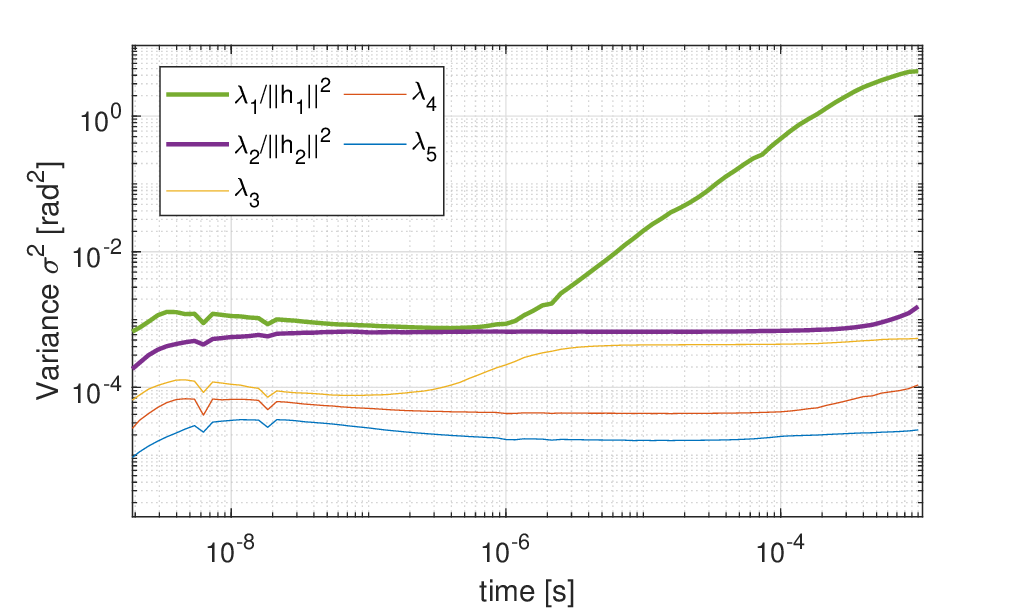}
\caption{(Experimental EO comb) Evolution of eigenvalues, of sample correlation matrix $\mathbf{S}(k)$, for $k=1,...,40\times 10^6 $, as a function of time $t = KT_s$.}
\label{fig:eigenvalues_vs_time_EO_exp}
\end{figure}

In Fig.~\ref{fig:eigenvalues_vs_time_EO_exp}, the evolution of the scaled eigenvalues is plotted as a function of time. We scale the first two eigenvalues, $\lambda_1$ and $\lambda_2$, according to \eqref{eq:lambda_i} to obtain variances of the phase noise sources, i.e.~$\sigma^2_{1_k}$ and $\sigma^2_{2_k}$. The remaining eigenvalues  $[\lambda_3,..,\lambda_5]$ represent variance of the residual measurement as noise shown in \eqref{eq:S_with_noise} and explain in section \ref{sec:num_results}, where we investigated the impact of measurement noise. 

The first thing to be observed is that the variance of the first phase noise sources, $\lambda_1 / \parallel \mathbf{h}_1 \parallel^2$, is the dominant source of phase noise. It starts dominating over the measurement noise for time exceeding $2\times 10^{-6}$ s. The second scaled eigenvalue (variance), $\lambda_2 / \parallel \mathbf{h}_2 \parallel^2$, is hard to distinguish from the measurement noise. It starts dominating over the measurement noise beyond  $5\times 10^{-5}$ s. This implies the second independent phase noise sources is lower than the measurement noise up to  $5\times 10^{-4}$ s.  \vspace{0.2cm}

\begin{figure}[h!]
  \centering
    \begin{subfigure}{0.49\linewidth}
        \includegraphics[width=\linewidth]{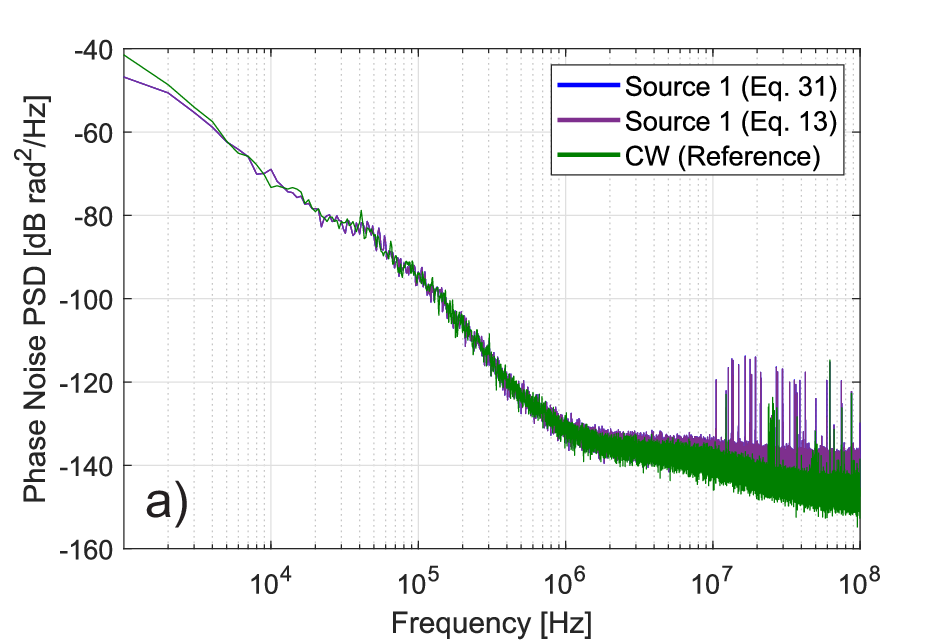}
    \end{subfigure}
    \begin{subfigure}{0.49\linewidth}
    \includegraphics[width=\linewidth]{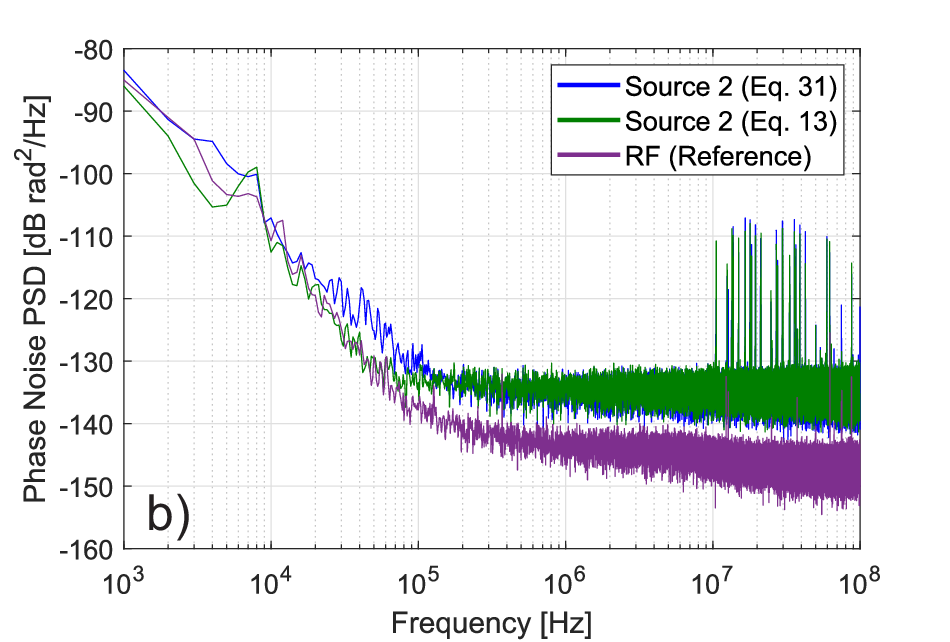}
    \end{subfigure}	
\caption{(Experimental EO comb) Power spectral density as a function of frequency for (a) the recovered first independent phase noise source and signal CW laser (b) the recovered second independent phase noise source and RF signal generator.}
\label{fig:PSD_EO_exp}
\end{figure}

In Fig.~\ref{fig:PSD_EO_exp}(a) and (b), the PSD of the recovered first and the second  phase noise sources associated with the first two eigenvectors/eigenvalues is plotted. For the reference, we also plot phase noise PSD of the CW signal laser and RF oscillator. The phase noise PSD of the CW signal laser is measured by removing the phase modulator in the setup shown in Fig.~\ref{setupeo}. To recover the phase noise of CW signal laser, we just use a single bandpass filter and perform the phase noise estimation as for the simulation results. The PSD of the RF oscillator is obtained by directly applying the RF oscillator at the ADC. The phase noise extraction then follows the same procedure as for the CW signal laser. 

Fig.~\ref{fig:PSD_EO_exp}(a), shows that the first phase noise source, (Source 1), obtained using \eqref{eq:norm_receovery_eq_Q_P} and \eqref{eq:matrixG_first_order}, respectively, agrees well with the the PSD of the CW signal laser source. This means that the first independent source of phase noise is equal to the phase noise of the CW laser used to EO comb generation. The CW signal laser phase noise will thus transfer equally to all comb-lines, exactly, as predicted theoretically, and confirmed numerically. 

In Fig.~\ref{fig:PSD_EO_exp}(b), we observe good agreement between the phase noise PSD of the second phase noise source, (Source 2), and the RF signal oscillator for frequencies below 100 kHz. For higher frequencies we observe higher measurement noise floor for the recovered independent phase noise source. This is attributed to the limited SNR of the EO comb lines, see Fig.~\ref{fig:spectrum_exp}, compared to the SNR for the direct phase noise measurement of RF oscillator. However, we can conclude the second phase noise source is equal to the phase noise of the RF oscillator. This implies that the phase noise contribution originating for the RF source will will increase linearly with com-line number. This is also in accordance with theoretical predictions and numerical investigations.


\subsection{Mode-locked laser}

\begin{figure*}[t!]
\centering \includegraphics[width=\textwidth]{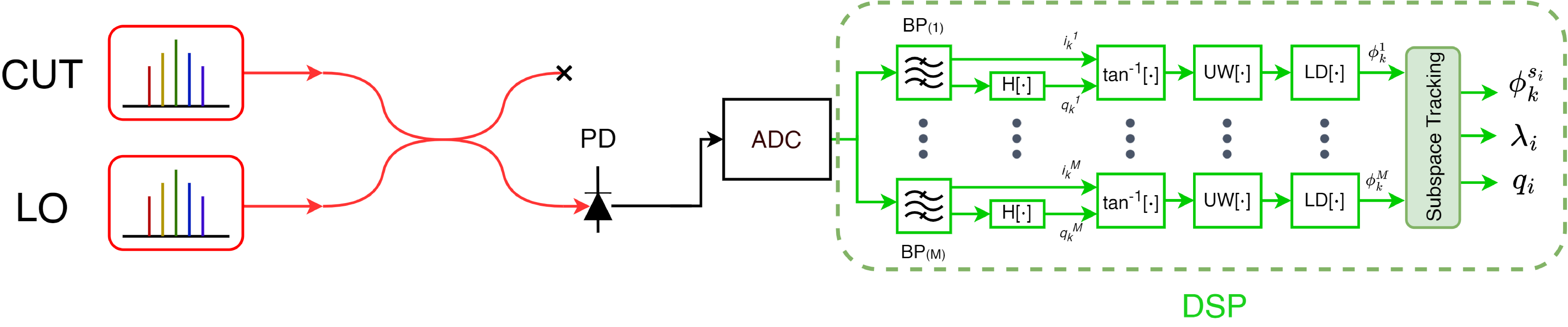}
\caption{Experimental setup for mode-locked laser phase noise characterization employing multi-heterodyne detection}
\label{fig:setupmll}
\end{figure*}

In this section, we proceed with the phase noise source identification and separation for a frequency modulated (FM) mode-locked laser (MLL). The experimental set-up is shown in Fig.~\ref{fig:setupmll}. The laser employed for the experiment is described in \cite{Dumont22}. The noise behavior of FM-MLL is less studied and understood compared to an EO comb. We can therefore use the proposed method based on subspace tracking to identify different phase noise sources according to the phase noise model depicted in (\eqref{eq:elastic tape model gen}). \vspace{0.2cm}

\begin{figure}[h!]
\centering \includegraphics[width=0.5\linewidth]{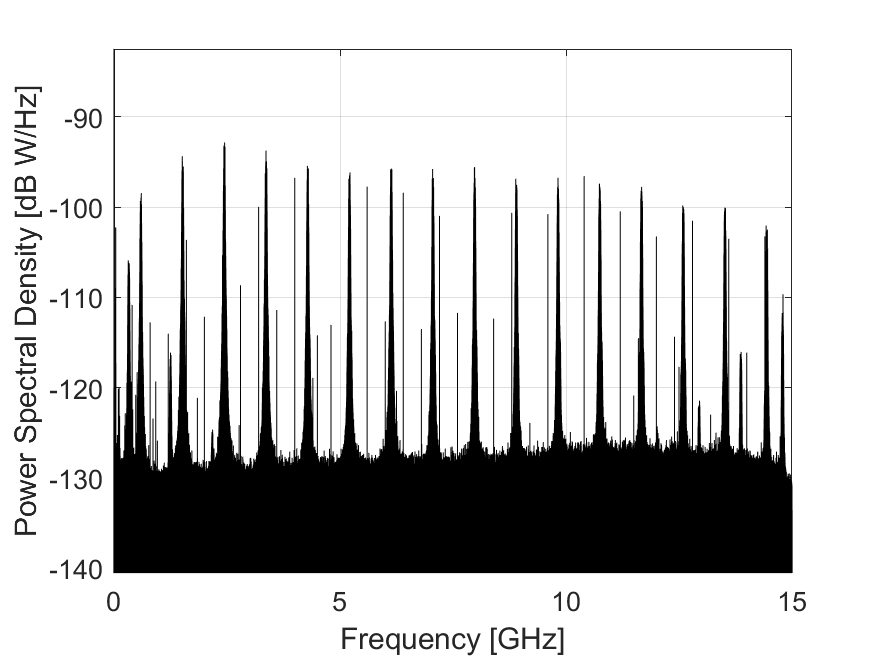}
\caption{(Mode-locked laser) Electrical spectrum of the downconverted MLL emission}
\label{fig:spectrum_MLL_exp}
\end{figure}

The employed MLLs  were designed in the same manner, with the same properties and conditions. We therefore believe that they have similar noise properties. They have 
though 1.7$\%$ difference in their free spectral ranges (FSRs). The mode-spacing of the two FM-MLLs are 58 and 59 GHz, both centered around 1285 nm with fiber coupled power of 9 mW each. When the combs are combined (with a 3 dB splitter), they are detected with a 15 GHz photoreceiver with transimpedance gain of 375 $\Omega$, which generates a cascade of heterodyne tones spaced by the difference in FSR, 1 GHz. A single-photodiode detection scheme was used in lieu of a balanced receiver to increase the strength of the weak individual beat notes (because each comb line has a power of $\approx$ 250 $ \mu$W). The beat signal was sampled with a real-time sampling scope having a bandwidth of 16 GHz and sampling rate of 32 GS/s. The electrical spectrum of the downconverted comb is shown in Fig.~\ref{fig:spectrum_MLL_exp}. For the phase noise analysis, we consider only $M=9$ down-converted comb-lines. The reason why we only consider 9 lines and not more is have sufficient signal-to-noise ratio.

\begin{figure}[h!]
\centering \includegraphics[width=0.7\linewidth]{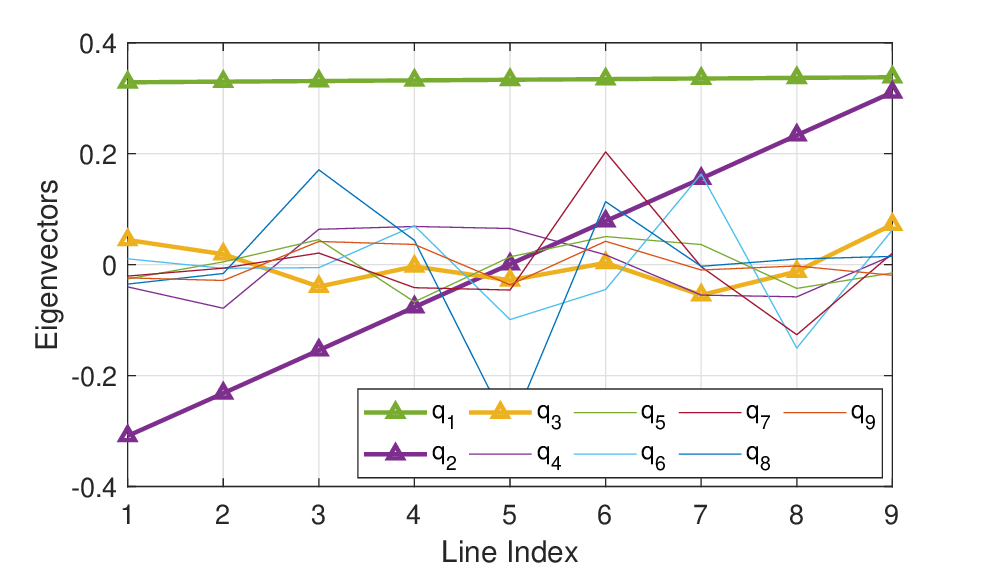}
\caption{(Mode-locked laser) Evolution of eigenvectors, of sample correlation matrix $\mathbf{S}(K)$, as a function of relative line index $m$ for $K=80\times10^6$.}
\label{fig:eigvectmll}
\end{figure}

\begin{figure}[h!]
\centering \includegraphics[width=0.8\linewidth]{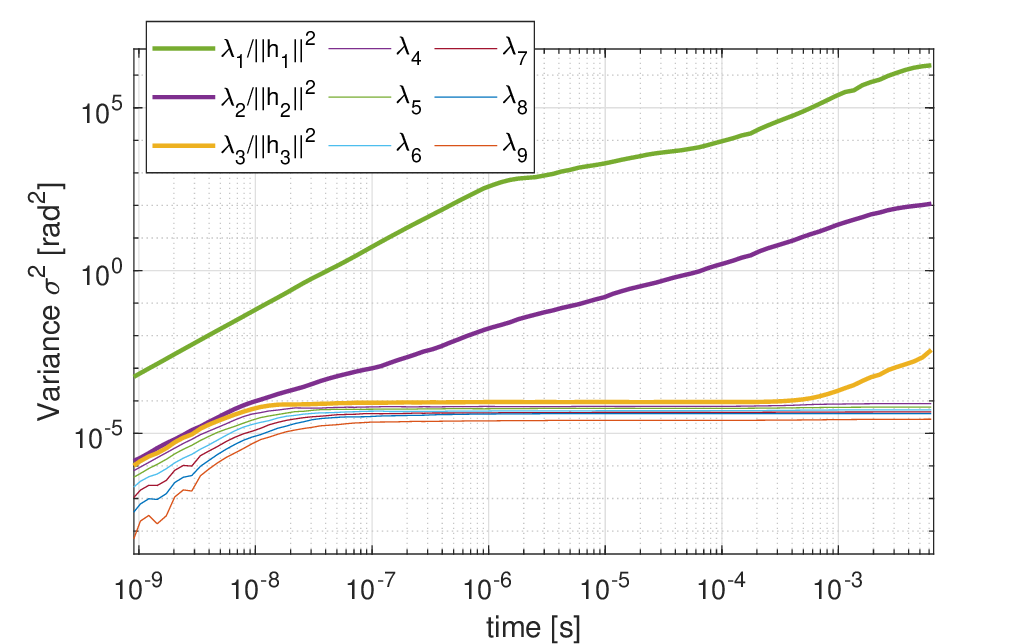}
\caption{(Mode-locked laser) Evolution of eigenvalues, of sample correlation matrix $\mathbf{S}(k)$, for $k=1,...,80\times 10^6 $, as a function of time $t = KT_s$.}
\label{fig:eigvalmll}
\end{figure}

\begin{figure}[h!]
\centering \includegraphics[width=0.7\linewidth]{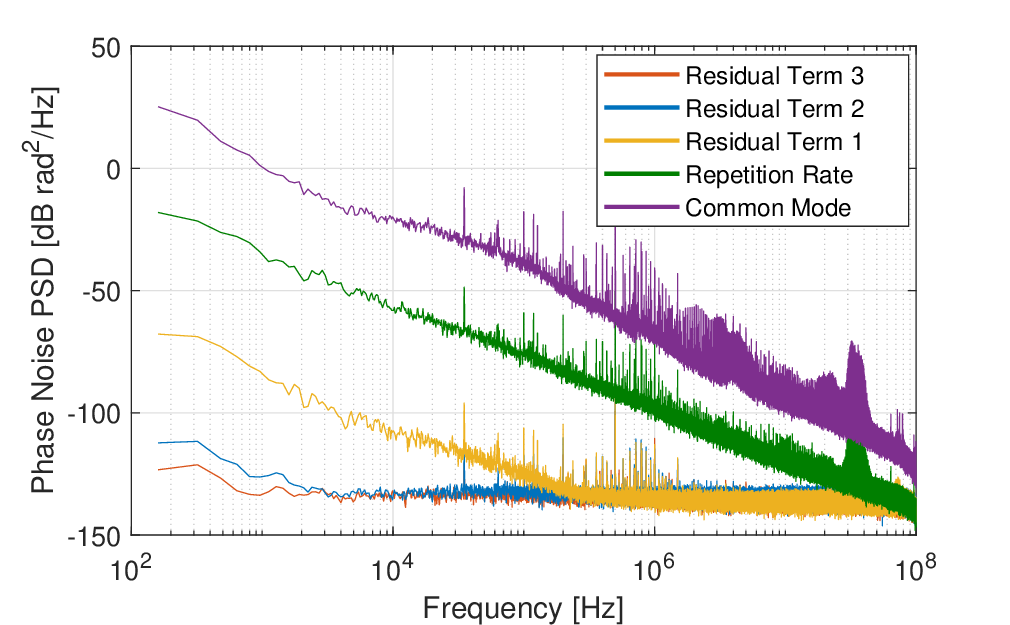}
\caption{(Mode-locked laser) Power spectral density as a function of frequency for the recovered first and the second independent phase noise source.}
\label{fig:psdmll}
\end{figure}

In Fig.~\ref{fig:eigvectmll}, we plot nine eigenvectors $[\mathbf{q}_1,...,\mathbf{q}_9]$. It is observed that the first eigenvector, $\mathbf{q}_1$, is constant as a function of comb-line number. The second eigenvector $\mathbf{q}_2$ increases linearly with the comb-line number. This is in accordance with the first two terms in Eq.~(\ref{eq:elastic tape model gen}). The remaining eigenvectors should represent the evolution of residual phase noise terms. Even though there is a large degree of fluctuation in eigenvectors $[\mathbf{q}_3,...,\mathbf{q}_9]$, the third eigenvector $\mathbf{q}_3$ seems to have most structure. However, it is difficult to draw any finite conclusions. We therefore plot the evolution of eigenvalues. \vspace{0.15cm}

In Fig.~\ref{fig:eigvalmll},  nine eigenvalues, $[\lambda_1,...,\lambda_9]$, are plotted as a function of time $KT_s$. It is clearly observed that the first two eigenvalues, $\lambda_1$ and $\lambda_2$ are distinguishable and increase with time. The first two eigenvalues are associated with common mode and repetition rate phase noise.  However, we also observe that the third eigenvalue $\lambda_3$ increases with time for $t>10^{-3}$ s, while the other remaining eigenvalues $[\lambda_4,...,\lambda_9]$ remain flat. If the eigenvalues are flat this implies that they represent measurement noise. On contrary since $\lambda_3$ is increasing with time, it represents residual phase noise. To get the full picture of the phase noise sources, in Fig.~\ref{fig:psdmll}, we show phase noise PSD of the recovered phase noise sources using \eqref{eq:norm_receovery_eq_Q_P}. \vspace{0.15cm}

The first two eigenvectors of matrix $\mathbf{Q}_P$ are used as a projection vectors to obtain phase noise PSDs of the common mode and the repetition rate phase noise, respectively. Since the first and the second eigenvector have constant and linear scaling, $||\mathbf{h}_1 ||$ and $||\mathbf{h}_2||$ of the normalization matrix $\mathbf{D}$ in Eq. ~(\ref{eq:normalization_matrix_D}) can be determined, i.e.~$||\mathbf{h}_1 ||=[1,...,1]$ and $||\mathbf{h}_2 ||=[-(M-1)/2,...,(M-1)/2]$ for $M=9$. Finding $||\mathbf{h}_3 ||$ to perform scaling is more difficult as the generation matrix is not known. To find $||\mathbf{h}_3 ||$, we use the prior knowledge that all phase noise PSDs must reach the same noise floor. This is a realistic assumption as the noise floor sets a lower limit for measurable phase noise PSDs. \vspace{0.15cm}

Finally, Fig.~\ref{fig:psdmll} clearly shows the presence of the residual phase noise source associated with the third eigenvector, i.e. ''Residual Term 1". The phase noise PSDs associated with the fourth and the fifth eigenvector, ''Residual Term 2" and ''Residual Term 3" have negligible magnitudes and are mostly flat, indicating that they may represent the noise floor.




\section{Conclusion} \label{sec:conc}

We introduce a novel method, based on subspace tracking and multi-heterodyne detection, for phase noise source separation in optical frequency combs.  The method can be used to separate phase noise sources associated with the common mode and repetition rate phase noise, as well as residual phase noise terms originating from technical noise or some other more complex sources of noise. This allows for computation of phase noise power spectrum densities associated with different noise source, as well as their scaling as a function of comb-line number. For our experimental study, we showed for the first time that apart from the common mode and repetition rate phase noise, the mode locked laser can also contain residual phase noise term.


\medskip
\noindent\textbf{Funding.} This work was supported by the European Research Council (ERC CoG FRECOM grant 771878), the Villum Foundations (VYI OPTIC-AI grant no. 29344) and DARPA MTO PIPES contract.

\noindent\textbf{Data availability.} Data underlying the results presented in this paper may be obtained from the authors upon reasonable request

\medskip
\noindent\textbf{Disclosures.} The authors declare no conflicts of interest.



\bibliography{bibliography}

\end{document}